\documentclass[12pt,a4paper]{article}

\usepackage[T1]{fontenc} 
\usepackage{microtype} 

\usepackage[hmarginratio=1:1,top=32mm,columnsep=20pt]{geometry} 

\usepackage{amsmath,amssymb}
\usepackage{hyperref}
\usepackage{cite}
\usepackage{graphicx}
\usepackage{caption}
\usepackage{lipsum}

\usepackage{abstract}

\usepackage{titlesec} 
\titleformat{\section}[block]{\large\bfseries\centering}{\thesection}{1em}{} 
\titleformat{\subsection}[block]{\bfseries}{\thesubsection}{1em}{} 
\numberwithin{equation}{section}

\usepackage{fancyhdr} 
\title{\vspace{-10mm}\fontsize{18pt}{22pt}\selectfont\textbf{Holography, Brane Intersections and Six-dimensional SCFTs }\vspace{3mm} }

\author{Nikolay Bobev$^1$, Giuseppe Dibitetto$^2$,\\ Fri\dh rik Freyr Gautason$^{1,3}$ and Brecht Truijen$^1$\\[5mm] 
\normalsize $^1$Instituut voor Theoretische Fysica, K.U. Leuven\\
\normalsize Celestijnenlaan 200D, BE-3001 Leuven, Belgium\\[1mm]
\normalsize $^2$Department of Physics and Astronomy, Uppsala University\\
\normalsize Box 516, SE-75120 Uppsala, Sweden\\[2mm]
\normalsize $^3$Institut de Physique Th\'eorique, Universit\'e Paris Saclay, CEA, CNRS,\\
\normalsize Orme des Merisiers, F-91191 Gif-sur-Yvette, France\\[3mm]
\texttt{\small\href{mailto:nikolay.bobev@kuleuven.be}{\{nikolay.bobev}, \href{mailto:ffg@kuleuven.be}{ffg}, \href{mailto:brecht.truijen@kuleuven.be}{brecht.truijen\}@kuleuven.be}}\\
\texttt{\small\href{mailto:giuseppe.dibitetto@physics.uu.se}{giuseppe.dibitetto@physics.uu.se}}\\ 
}
\date{}

\usepackage{amsthm}

\usepackage{enumerate}
\numberwithin{equation}{section}

\usepackage{color}
\definecolor{dark-gray}{gray}{0.20}
\definecolor{gray}{gray}{0.30}
\definecolor{light-gray}{gray}{0.80}
\definecolor{dark-red}{rgb}{0.7,0,0}
\definecolor{dark-green}{rgb}{0.1,0.4,0}
\definecolor{dark-blue}{rgb}{0.3,0.3,0.7}
\definecolor{light-blue}{rgb}{0.8,0.8,1}

\usepackage{cite}
\usepackage{hyperref}
\hypersetup{
	colorlinks=true,
	linkcolor=dark-blue,
	citecolor=dark-red,
	urlcolor=dark-green,
	linktoc=page
}

\newcommand{\dd}{\mathrm{d}}

\newcommand{\e}{\mathrm{e}}
\newcommand{\w}{\wedge}

\newcommand{\be}{\begin{equation}}
\newcommand{\ee}{\end{equation}}
\newcommand{\bea}{\begin{eqnarray}}
\newcommand{\eea}{\end{eqnarray}}

\newcommand{\f}[2]{\frac{#1}{#2}}

\newcommand{\vol}{\text{vol}}

\definecolor{cardinal}{rgb}{0.6,0,0}
\definecolor{darkgreen}{rgb}{0,0.5,0}
\definecolor{golden}{rgb}{0.92, 0.7, 0}
\definecolor{midnight}{rgb}{0, 0, 0.5}
\definecolor{darkblue}{rgb}{0.2, 0, 0.8}

\def\be{\begin{equation}}
\def\ee{\end{equation}}

\begin{document}

\begin{flushright}
\small UUITP-32/16\\
\normalsize
\end{flushright}

{\let\newpage\relax\maketitle}
\thispagestyle{empty}

\begin{abstract}
We study supersymmetric intersections of NS5-, D6- and D8-branes in type IIA string theory. We focus on the supergravity description of this system and identify a ``near horizon'' limit in which we recover the recently classified supersymmetric seven--dimensional AdS solutions of massive type IIA supergravity. Using a consistent truncation to seven-dimensional gauged supergravity we construct a universal supersymmetric deformation of these AdS vacua. In the holographic dual six-dimensional (1,0) superconformal field theory this deformation describes a universal RG flow on the tensor branch of the vacuum moduli space triggered by a vacuum expectation value for a protected scalar operator of dimension four.

\end{abstract}

\newpage

\setcounter{tocdepth}{2}
\pagenumbering{arabic}
\tableofcontents

\section{Introduction}

Six-dimensional interacting SCFTs provide an interesting and exotic corner of the landscape of consistent QFTs. Early hints for their existence came from studying the low-energy dynamics of brane intersections in string and M-theory \cite{Seiberg:1996qx,Brunner:1997gf,Hanany:1997gh}. It is believed that the list of six-dimensional  $\mathcal{N}=(2,0)$ SCFTs is exhausted by the theories labeled by the $ADE$ algebras. The kaleidoscope of $\mathcal{N}=(1,0)$ SCFTs appears to be much richer and a full classification of such theories is still lacking. Recently there has been a revival in this area sparked by advances in F-theory \cite{Heckman:2013pva,DelZotto:2014hpa,Heckman:2015bfa} and holographic constructions \cite{Gaiotto:2014lca,Apruzzi:2013yva,Apruzzi:2015zna,Rota:2015aoa,Apruzzi:2015wna},  as well as our better understanding of the anomaly polynomials of six-dimensional supersymmetric theories \cite{Ohmori:2014kda}.\footnote{See \cite{Tomasiello:2016wfy} for a review on these recent developments with a more exhaustive list of references, and \cite{Bhardwaj:2015xxa} for a Lagrangian-based approach to classifying anomaly-free six-dimensional supersymmetric QFTs.} This renewed interest is well justified, since understanding the structure of six-dimensional interacting CFTs is bound to teach us important lessons about the mysterious theory living on the world-volume of M5-branes. In addition, compactifications of six-dimensional theories lead to new insights into the physics of lower-dimensional QFTs and the dualities that they enjoy.

Our interest here is in the class of linear quiver six-dimensional SCFTs introduced in \cite{Brunner:1997gf,Hanany:1997gh} and explored recently with new tools by Gaiotto and Tomasiello \cite{Gaiotto:2014lca}. In field theory language this setup is the six-dimensional analog of the usual Hanany-Witten type linear quivers which are well-studied in the context of three-dimensional $\mathcal{N}=4$ \cite{Hanany:1996ie} and four-dimensional $\mathcal{N}=2$ \cite{Witten:1997sc} theories. One starts with a particular brane intersection of NS5-, D6-, and D8-branes in type IIA string theory in which the branes share five flat spatial and one temporal direction.\footnote{This is summarized in Table \ref{table:intersection} below.} When the NS5-branes are separated along the worldvolume of the D6-branes one has a description of the low-energy theory as a six-dimensional quiver gauge theory. Each segment of $n$ D6-branes leads to an $SU(n)$ gauge group. The D8-branes transverse to each segment add ``flavor'' hyper multiplets in the fundamental of the gauge group, while the NS5-branes cary bi-fundamental hyper multiplets. The relative separation between the NS5-branes is controlled by the real scalar in a six-dimensional tensor multiplet. When the vacuum expectation value for this scalar vanishes the NS5-branes coincide and one finds an interacting $\mathcal{N}=(1,0)$ SCFT. It was argued in \cite{Gaiotto:2014lca} that when the number of NS5-branes is large these SCFTs admit a dual holographic description in terms of type IIA supergravity on the $AdS_7$ backgrounds classified and studied in \cite{Apruzzi:2013yva}. These $AdS_7$ solutions are constructed directly in type IIA supergravity without any direct reference to the underlying brane construction \cite{Apruzzi:2013yva}. While there is substantial evidence for the validity of the holographic duality proposed in \cite{Gaiotto:2014lca} (see for example \cite{Cremonesi:2015bld}) we believe that there is room for improvement. 

The ``gold standard'' of the AdS/CFT correspondence is the duality between type IIB  string theory on the $AdS_5\times S^5$ background and the $\mathcal{N}=4$ SYM theory \cite{Maldacena:1997re}. The key to understanding this duality is provided by the underlying D3-branes. To obtain the $AdS_5\times S^5$ solution of type IIB string theory in the supergravity limit one starts from the asymptotically flat space solution describing $N$ coincident D3-branes, which in turn can be thought of as an extremal black brane. Then one takes an appropriate near-horizon limit to isolate the $AdS_5\times S^5$ region. The same procedure can be applied to D3-branes at singular CY three-folds and it leads to a plethora of $AdS_5/CFT_4$ holographically dual pairs. The $AdS_7/CFT_6$ duality studied in \cite{Gaiotto:2014lca} is on a different footing. The reason is that the $AdS_7$ solutions of \cite{Apruzzi:2013yva} have not been shown to arise from some type of near-horizon limit of intersecting brane solutions in massive type IIA supergravity. The goal of our work is to fill in this gap.

Our starting point is a careful analysis of the system of BPS equations derived by Imamura in \cite{Imamura:2001cr}. These equations control supersymmetric solutions of massive type IIA supergravity which should describe the backreaction of a system of intersecting NS5-, D6-, and D8-branes. Finding solutions to these non-linear partial differential equations in general is a non-trivial problem. We make progress using several different approaches. First, we impose an Ansatz for all background fields of type IIA supergravity which is invariant under the isometries of $AdS_7$. Upon a judicious choice of coordinates this leads to a drastic simplification and the BPS equations reduce to a simple system of coupled ordinary differential equations which we solve explicitly. In this way we recover the supersymmetric $AdS_7$ solutions classified in \cite{Apruzzi:2013yva}. Equipped with these explicit solutions we then proceed to study deformations which break the isometries of $AdS_7$ and are holographically dual to supersymmetric RG flows in the $\mathcal{N}=(1,0)$ SCFTs of \cite{Gaiotto:2014lca}. An important technical ingredient in our analysis is the existence of a consistent truncation of massive type IIA supergravity to minimal seven-dimensional gauged-supergravity established in \cite{Passias:2015gya}. The holographic RG flows of interest are particularly simple analytic solutions of this seven-dimensional supergravity which can be readily uplifted to ten or eleven dimensions. The uplifted backgrounds in turn provide nontrivial examples of explicit analytic solutions to the non-linear PDEs of \cite{Imamura:2001cr}. These backgrounds can be interpreted as sourced by smeared NS5-branes in type IIA supergravity with a particular charge density controlled by the conformal symmetry breaking parameter in the dual RG flow. Equipped with some intuition from these explicit solutions we are able also to construct more general supersymmetric backgrounds in type IIA supergravity with vanishing Romans mass. They correspond to a general charge distribution of NS5-branes along a stack of D6-branes.

In addition to understanding how the $AdS_7$ solutions of \cite{Apruzzi:2013yva} arise as the particular brane intersections suggested by the field theory construction of \cite{Gaiotto:2014lca} a further motivation for our work is to study supersymmetric deformations of these six-dimensional SCFTs using holography. The deformations of $AdS_7$ mentioned above, correspond to supersymmetric RG flows in the dual SCFT triggered by a dimension four scalar operator. This operator is the lowest component in the energy-momentum tensor multiplet and is thus present in every $\mathcal{N}=(1,0)$ SCFT. In harmony with the results in \cite{Cordova:2016xhm} we find that the only possible supersymmetric and Lorentz-invariant deformation of the $\mathcal{N}=(1,0)$ SCFT at hand is realized by turning on a vacuum expectation value (vev) for this operator. This vev parametrizes a particular direction in the tensor branch of the $\mathcal{N}=(1,0)$ SCFT. Our holographic construction suggests that such RG flows on the tensor branch, at least in some appropriate large $N$ limit, have a universal nature which is independent of the details of the six-dimensional theory.

We start our exploration in the next section by reviewing the salient features of the 6d $\mathcal{N}=(1,0)$ SCFTs arising from intersecting D6-, NS5-, and D8-branes in type IIA string theory. In Section \ref{sec:Imamura} we switch gears to supergravity to discuss the intersecting brane Ansatz and BPS equations of \cite{Imamura:2001cr} and show how the $AdS_7$ solutions of \cite{Apruzzi:2013yva,Rota:2015aoa} arise as solutions of these equations. Section \ref{sec:7dflows} is devoted to a construction of an explicit supergravity solution which is holographically dual to a particular tensor branch deformation in the six-dimensional SCFTs. In Section \ref{sec:masslessflatsol} we discuss a new type IIA supergravity solution which describes an intersection of NS5- and D6-branes and relate it to the discussion in Section \ref{sec:7dflows}. We conclude with a brief summary of our results and possible directions for future study in Section \ref{sec:conclusions}. The three appendices contain our conventions, some details on the derivation of the $AdS_7$ solutions of interest, and an explicit relation between the BPS equations derived in \cite{Imamura:2001cr} and those of \cite{Apruzzi:2013yva,Rota:2015aoa}.

\paragraph{Note added:} After the submission of this manuscript to the arXiv we became aware of the work in \cite{Macpherson:2016xwk} which has partial overlap with our results in Section \ref{sec:Imamura}.

\section{Brane intersections and six-dimensional SCFTs}
\label{sec:6dSCFTs}

We are interested in six-dimensional $\mathcal{N}=(1,0)$ supersymmetric QFTs. These theories preserve eight real chiral supercharges and the R-symmetry group is $SU(2)$. The supersymmetric multiplets are the usual vector and hyper multiplets familiar from theories with eight supercharges in three and four dimensions, as well as the more exotic tensor multiplet. The only bosonic field in the vector multiplet is the gauge field $A_{\mu}$ with field strength $F_{\mu\nu}$. Therefore, in contrast to three and four-dimensional supersymmetric theories, there is no Coulomb branch of the vacuum moduli space since there are no scalars in the vector multiplet. In the hyper multiplet we have four real scalars. These parametrize the Higgs branch which has a structure similar to the one of four-dimensional $\mathcal{N}=2$ theories. The tensor multiplet contains one real scalar field, $\phi$, and a two-form tensor potential, $b_{\mu\nu}$, with a self-dual field strength, $h_{\mu\nu\rho}$. The vacuum expectation value of the scalar, $\phi$, in the tensor multiplet parametrizes a branch of the vacuum moduli space called the tensor branch. This will play an important role in our story. To illustrate how this works schematically we present the relevant terms of the bosonic Lagrangian for an Abelian tensor multiplet coupled to a gauge field
\begin{equation}\label{eq:6dLag}
\mathcal{L} \supset \phi \text{Tr}\left(F_{\mu\nu}F^{\mu\nu}\right) +\partial_{\mu}\phi\partial^{\mu}\phi + h_{\mu\nu\rho}h^{\mu\nu\rho} + \star \left(b\wedge \text{Tr}\left(F\wedge F\right)\right)\;.
\end{equation}
The operator $\phi$ is gauge invariant and its classical scaling dimension is 2. Its vacuum expectation value, $\langle\phi\rangle$ parametrizes the tensor branch of the moduli space. Here we have restricted ourselves to one tensor multiplet for simplicity. The  vev $\langle\phi\rangle$ can be thought of as the effective gauge coupling $\langle\phi\rangle \sim 1/g_{YM}^2$ and the singular point $\langle\phi\rangle=0$ should be analyzed with care. Crucial insight from string theory suggests that the limit $\langle\phi\rangle \to 0$ often corresponds to a critical point of the renormalization group flow and thus an interacting SCFT \cite{Seiberg:1996qx}. In fact to the best of our knowledge all known examples of interacting six-dimensional CFTs are supersymmetric and arise from suitable constructions in string, M-, or F-theory.

\begin{figure}[h!]
\begin{center}
\includegraphics[width=1.0\textwidth]{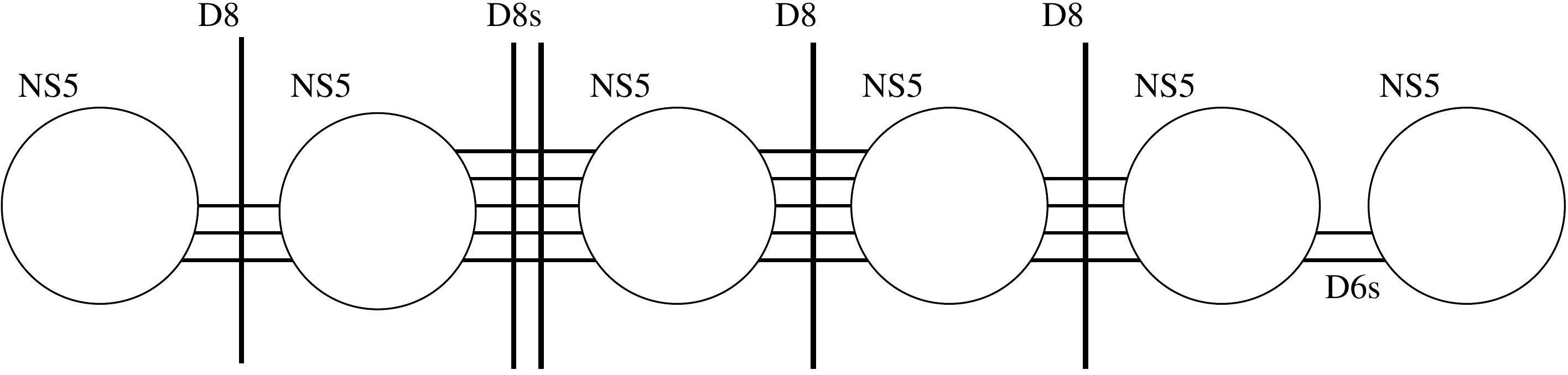}
\caption{\label{fig:separatedNSbranes}  An illustrative example of the system of intersecting branes discussed in the main text and in Table \ref{table:intersection}.} 
\end{center}
\end{figure}

\begin{figure}[h!]
\begin{center}
\includegraphics[width=1.0\textwidth]{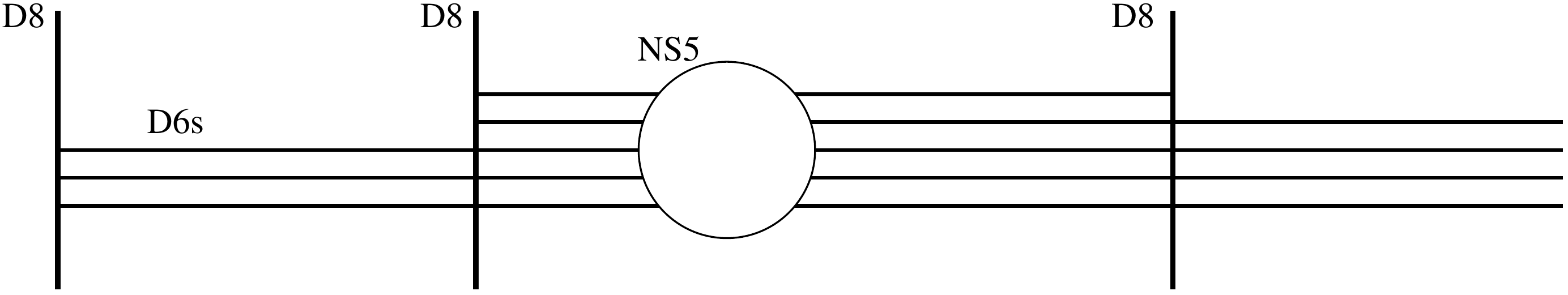}
\caption{\label{fig:branes}  A brane configuration that should be described by the conformal limit of a linear quiver gauge theory.} 
\end{center}
\end{figure}

The six-dimensional supersymmetric theories of interest to us are the linear quivers introduced in \cite{Brunner:1997gf,Hanany:1997gh} and further studied in \cite{Gaiotto:2014lca}. These are six-dimensional cousins of the three- and four-dimensional linear quiver gauge theories with eight supercharges \cite{Hanany:1996ie,Witten:1997sc}. The six-dimensional gauge theories describe the low-energy dynamics of a system of NS5-, D6-, and D8-branes in flat space arranged according to the diagram in Table \ref{table:intersection}.\footnote{One could also introduce appropriate orientifold planes in this construction while still preserving $\mathcal{N}=(1,0)$ supersymmetry. See \cite{Gaiotto:2014lca} for more details.} The six-dimensional vector multiplets of the gauge theory arise from the worldvolume dynamics of the D6-branes. The gauge group is $SU(n)$ for a segment of $n$ D6-branes in the $z$  direction. The D8-branes intersect the D6-branes at isolated points on the $z$ line and lead to hypermultiplets in the fundamental representation of the gauge group. The NS5-branes are point-like on the line parametrized by $z$. Each NS5-brane leads to a bi-fundamental hypermultiplet associated with the two stacks of D6-branes that end on the given NS5-brane. In addition each pair of NS5-branes contains a tensor multiplet. The vev for the real scalar field in this multiplet corresponds to the distance between the NS5-branes in the $z$ direction. In general there are many such NS5-branes with generic values of these real vevs. This situation corresponds to a general point on the tensor branch of the six-dimensional theory and is illustrated by the diagram in Figure \ref{fig:separatedNSbranes}. When the NS5-branes coincide all the tensor multiplet scalars have a vanishing vev and one is at the origin of the tensor branch where it is expect that a strongly interacting SCFT resides.\footnote{In the absence of D8-branes these six-dimensional theories are the same as the $\mathcal{N}=(1,0)$ theories of type $(A_{N},A_{k})$ obtained by placing $N$ M5-branes on a $\mathbb{Z}_k$ singularity in M-theory. Here $N$ and $k$ are the numbers of NS5- and D6- branes, respectively.} This is illustrated by the diagram in Figure \ref{fig:branes}. These SCFTs are strongly coupled and evade a Lagrangian description. In a suitable limit when the number of coinciding NS5-branes is large it was argued in \cite{Gaiotto:2014lca} that these SCFTs are dual to the supersymmetric $AdS_7$ solutions of massive type IIA supergravity found in \cite{Apruzzi:2013yva}.

\begin{table}[h!]
\begin{center}
\begin{tabular}{ccccccccccc}
\hline
\hline
	&$t$	&$x^1$	&$x^2$	&$x^3$	&$x^4$	&$x^5$	&$z$	&$r$	&$\theta$	&$\phi$\rule{0pt}{2.5ex}\\
\hline
NS5	&$\circ$&$\circ$&$\circ$&$\circ$&$\circ$&$\circ$&		&		&			&		\\
D6	&$\circ$&$\circ$&$\circ$&$\circ$&$\circ$&$\circ$&$\circ$&		&			&		\\
D8	&$\circ$&$\circ$&$\circ$&$\circ$&$\circ$&$\circ$&		&$\circ$&$\circ$	&$\circ$\\
\hline
\end{tabular}
\caption{\label{table:intersection}The brane intersection in type IIA string theory that leads to the SCFTs and supergravity solutions of interest in this work.}
\end{center}
\end{table}

In the absence of a Lagrangian it is often instructive to adopt an algebraic approach to study SCFTs. Every six-dimensional $\mathcal{N}=(1,0)$ SCFT should contain an energy-momentum tensor which belongs to a particular short multiplet of the $OSp(8|2)$ superconformal algebra. The bosonic content of the energy-momentum tensor multiplet is:\footnote{See for example Table 31 in \cite{Cordova:2016xhm}.} a scalar operator, $\mathcal{O}$, of conformal dimensions 4 which is neutral under the R-symmetry; the $SU(2)$ R-current, $\mathcal{J}_{\mu}$, which has conformal dimension 5 and is in the spin-1 representation of $SU(2)$; another operator of dimension 5, $\mathcal{S}^{+}_{\mu\nu\rho}$, which transforms as a self-dual 3-form under the six-dimensional Lorentz group and is neutral under the R-symmetry; and the energy momentum tensor, $\mathcal{T}_{\mu\nu}$, which is a symmetric rank two tensor of conformal dimension 6 and is neutral under the R-symmetry. It was shown in \cite{Cordova:2016xhm} using superconformal algebraic methods that there are no supersymmetric, Lorentz-invariant, relevant or marginal deformations of $\mathcal{N}=(1,0)$ superconformal theories (see also \cite{Louis:2015mka,Buican:2016hpb}). Thus the only possible Lorentz invariant supersymmetric RG flows in such SCFTs are obtained by vevs, i.e. by moving on the vacuum moduli space. This moduli space consists of two branches - the tensor branch where the $SU(2)_R$ symmetry is unbroken and the Higgs branch where it is broken. For a recent review and references to the original literature see \cite{Cordova:2015fha}. All known six-dimensional interacting SCFTs have a tensor branch. This state of affairs is similar to the situation in four-dimensional interacting $\mathcal{N}=2$ SCFTs  which all appear to have a Coulomb branch. In general the tensor branch is multi-dimensional. For example in the linear quiver gauge theories discussed above each pair of NS5-branes carries a tensor multiplet and thus adds one real dimension to the tensor branch. In anticipation of the supergravity results in Section \ref{sec:7dflows} we should point out that the holographic RG flows discussed there describe some particular direction in this multi-dimensional tensor branch. This direction is singled out since it is parametrized by the vev for the dimension 4 scalar operator $\mathcal{O}$ discussed above.

After this short foray into the world of six-dimensional theories with  $\mathcal{N}=(1,0)$ supersymmetry it is time to move to a more detailed discussion of their dual supergravity description.

\section{Supergravity description}
\label{sec:Imamura}

Our goal is to construct supersymmetric solutions of massive type IIA supergravity \cite{Romans:1985tz} which describe the backreaction of the system of intersecting NS5-, D6- and D8-branes presented in Table \ref{table:intersection}. This problem was addressed by Imamura in \cite{Imamura:2001cr} and below we will heavily exploit his results. Starting from the brane intersection in Table \ref{table:intersection}, we impose Poincar\'e invariance along the shared worldvolume of the branes spanned by $t,x^1,\dots,x^5$ and unbroken $SO(3)$ isometry along a two-sphere parametrized by the angles $\theta$ and $\phi$. All background fields in the supergravity theory are in general non-trivial functions of the coordinates $r$ and $z$. Type IIA supergravity has a number of form fields which are also assumed to respect the Poincar\'e symmetry and $SO(3)$ isometry of the metric. These are the RR 2-form $F_2$ which has legs along $\theta$ and $\phi$ and the NSNS 3-form $H$ which has both $r\theta\phi$ and $z\theta\phi$ components. It was argued in \cite{Imamura:2001cr} that both the $rz$ component of $F_2$ and the entire RR 4-form $F_4$  vanish. Finally in order to preserve 1/4 of the maximal supersymmetry one has to impose that the supersymmetry variations of the type IIA gravitino and dilatino vanish subject to the following projection conditions
\be\label{projectors}
\epsilon^2 = \Gamma_{r\theta\phi}\epsilon^1~,\qquad \epsilon^2 = \Gamma_{z}\epsilon^1~.
\ee
Here $\epsilon^{1,2}$ are 16-component Majorana-Weyl spinors. The first relation in \eqref{projectors} is the familiar spinor projection satisfied by the supersymmetry parameter of D6-branes in flat space, whereas combining the two equations in \eqref{projectors} gives the analogous spinor projector for NS5-branes. The resulting BPS equations can be solved and lead to the field configuration\footnote{We work in string frame. Our supergravity conventions can be found in Appendix \ref{notation}. }
\bea
\dd s^2 &=& S^{-1/2} \dd s_6^2 + K\left[ S^{-1/2} \dd z^2 + S^{1/2} (\dd r^2 + r^2 \dd \Omega_2^2)\right]~,\label{ansatz1}\\
\e^{2\phi} &=& g_s^2 K S^{-3/2}~,\label{ansatz2}\\
F_2 &=& - r^2 g_s^{-1} \partial_r S\; \vol_2~,\label{ansatz3}\\
H &=& -r^2 \left[\partial_r K \dd z - \partial_z(KS)\dd r\right]\w \vol_2~,\label{ansatz4}
\eea
where $\dd s_6^2$ is the flat metric on six-dimensional Minkowski space, $\dd \Omega_2^2$ and $\vol_2$ are the Einstein metric and the volume form on the round two-sphere $S^2$.\footnote{It is compatible with supersymmetry to replace $S^2$ with $\mathbb{RP}^2$. We thus have this freedom for all supergravity solutions discussed below.} The functions $S$ and $K$ depend on $r$ and $z$ and can be thought of as the ``harmonic functions'' associated with D6- and NS5-branes respectively. The Bianchi identities for $F_2$ and $H$,
\be
\dd F_2 - M H =0\;, \qquad \dd H = 0~,
\ee
imply three partial differential equations for $S$ and $K$: 
\begin{eqnarray}
\partial_z S - M g_s K &=& 0 ~,\nonumber\\
\triangle_3 S + M g_s \partial_z(KS) &=& 0~,\label{imamuraeq}\\
\triangle_3 K + \partial_z^2(KS) &=& 0~,\nonumber
\end{eqnarray}
where $\triangle_3 = r^{-2}\partial_r r^2 \partial_r$. For non-vanishing Romans mass, $M\neq 0$, this system can be rewritten as a single non-linear equation for the function $S$
\begin{equation}\label{eq:1Seq}
\triangle_3 S  +\frac{1}{2}\partial_z^2S^2 = 0\;.
\end{equation}
Given a solution to this equation the function $K$ is then determined through the first equation in \eqref{imamuraeq}. For $M=0$ one finds that $\partial_zS=0$ and the last two equations in  \eqref{imamuraeq} have to be solved as a coupled system.

The system of equations in \eqref{imamuraeq} is in general non-linear which is a well-known feature of the BPS equations controlling brane solutions of massive type IIA supergravity (see for example \cite{Janssen:1999sa}). In the limit of vanishing Romans mass, $M=0$, the system in  \eqref{imamuraeq} becomes linear and should describe backreacted NS5-D6-brane solutions. At this point it is worth presenting some simple well-known solutions of IIA supergravity in the massless limit that fit into this general discussion: 

\begin{itemize}

\item The solution corresponding to a stack of $N_6$ D6 branes localized at $r=0$ is given by
\begin{equation}\label{D6branesol}
M=0\;, \qquad K=1\;, \qquad S = 1+\frac{N_6 g_s}{4\pi r}\;.
\end{equation}
\item The solution corresponding to a stack of $N_5$ NS5-branes localized at $z=r=0$ is given by
\begin{equation}\label{NS5branesol}
M=0\;, \qquad K=1+\frac{N_5 g_s}{4\pi^2 (r^2+z^2)}\;, \qquad S = 1\;.
\end{equation}
\item The solution corresponding to a stack of $N_6$ branes localized at $r=0$ and NS5 branes smeared along $z$ with density $\rho_5$ is
\begin{equation}
M=0\;, \qquad K=1+\frac{\rho_5}{4\pi r}\;, \qquad S = 1+\frac{N_6 g_s}{4\pi r}\;.
\end{equation}

\end{itemize}

Thus in the massless limit of type IIA supergravity the function $K$ can be thought of as the harmonic function associated with the NS5-branes and the function $S$ the one associated with the D6-branes.

\subsection{$AdS_7$ solutions}
\label{subsec:AdS7}

As reviewed in Section \ref{sec:6dSCFTs} one can obtain interacting six-dimensional $(1,0)$ SCFTs from the intersection of NS5-, D6- and D8-branes in type IIA string theory summarized in Table \ref{table:intersection}. It is thus natural to expect that the system of BPS equations \eqref{imamuraeq} admits $AdS_7$ solutions which provide a dual holographic description of these interacting SCFTs. In this section we determine the conditions on the functions $K(r,z)$ and $S(r,z)$ under which the system of equations \eqref{imamuraeq} leads to $AdS_7$ solutions.

The strategy is to combine the coordinates $z$ and $r$ to form the radial coordinate of $AdS_7$ which we call $\rho$. We use the following parametrization of the metric on $AdS_7$:
\be
\dd s_7^2 = \f{1}{(g\rho)^2}\dd\rho^2 + (g\rho)\dd s_6^2~,
\ee
where $\dd s_6^2$ is the flat Minkowski metric as in \eqref{ansatz1}, and $g$ is related to the AdS radius $L$ through $L = 2/g$. The other independent combination of $r$ and $z$ will form a coordinate which we call $\alpha$. This coordinate, combined with the coordinates on the two-sphere $\dd \Omega_2^2$ in \eqref{ansatz1}, forms a three-dimensional space $\mathcal{M}_3$. Upon finding an explicit solution for the metric one then has to properly analyze the global properties of $\mathcal{M}_3$ in order to understand the physics of the $AdS_7$ solution.

In Appendix \ref{app:AdS7} we summarize the analysis of equations \eqref{imamuraeq} which ensures that the background fields of type IIA supergravity in \eqref{ansatz1}-\eqref{ansatz4} obey the isometries of $AdS_7$. The upshot is that one finds the following relation between the radial variable of AdS $\rho$ and the coordinates $(r,z)$:
\be
\rho^{-1} = g^3(z^2 + 4r^2 S)K~.
\ee
In addition one finds that the functions $S$ and $K$ must satisfy the following differential constraints:
\begin{equation}\label{thirdeq}
\begin{split}
2S + 2r\partial_rS + z\partial_z S &= 0~,\\
3K + 2r\partial_rK + z\partial_z K &= 0~,\\
-z\partial_r K + 2r\partial_z(KS) &=0~.
\end{split}
\end{equation}
The first two equations in \eqref{thirdeq} can be integrated to give
\be
K = \frac{2}{z^3} G(r/z^2)~,\qquad\qquad S = \frac{1}{2g^2r}y(r/z^2)~,\label{SKyGeq}
\ee
where $y$ and $G$ are so far undetermined functions of the variable $r/z^2$. One can then show that the internal coordinate $\alpha$ is also a function of $r/z^2$. It proves convenient to use the following parametrization of $\alpha$:
\be\label{alphacoordinate}
\alpha \equiv 4 g^2\left(g^2+2\f{r}{z^2}y(r/z^2)\right) G(r/z^2) = \f{2gz}{\rho}~,
\ee
In addition it is beneficial to define the following function of $r/z^2$ 
\be\label{eq:betadef}
\beta(\alpha)\equiv \f{r}{2g^2z^2} \alpha(r/z^2)^2~.
\ee
It is important to emphasize that since $\alpha$ depends only on $r/z^2$ from now on we will consider $\beta$ and $y$ to be a function solely of $\alpha$. After all of these coordinate changes and redefinitions one can show that the system of BPS equations in \eqref{imamuraeq} together with the constraints in \eqref{thirdeq} reduce to the following pair of simple ODEs:
\begin{equation}\label{AdsODE}
\begin{split}
2y(\alpha)y'(\alpha) - M g_s  =& 0 ~,\\ 
2y(\alpha) \beta'(\alpha) + \alpha =& 0~,
\end{split}
\end{equation}
where the prime denotes a derivative with respect to $\alpha$. Our analysis so far has shown that a solution to the equations in \eqref{AdsODE} together with the definitions in \eqref{SKyGeq}, \eqref{alphacoordinate}, and \eqref{eq:betadef} leads to an $AdS_7$ solution to the system of BPS equations in \eqref{imamuraeq}. In fact, the metric and background fields of type IIA supergravity can now be written explicitly in terms of $\alpha$, $y$, and $\beta$:
\begin{equation}\label{metricfinal}
\begin{split}
\dd s^2 =& \sqrt{\f{\beta}{y}}\left(\dd s_7^2 + \f{1}{g^2 \beta y}\left(\f{\dd \alpha^2}{4} + 
\f{(\beta y)^2}{\alpha^2 + 4 y \beta}\dd\Omega_2^2\right)\right)~,\\
\e^{4\phi} =& \f{16 g^4 g_s^4 \beta^3}{y^3(\alpha^2 + 4y\beta)^2}~,\\
F_2 =& \f{1}{2g^2 g_s}\left( y + \f{M g_s\beta\alpha}{\alpha^2 + 4y\beta}\right)\vol_2~,\\
H =& \f{\beta}{2g^2y(\alpha^2 + 4y\beta)}\left(3y - \f{2M g_s\beta\alpha}{\alpha^2 + 4y\beta}\right)\dd\alpha\w\vol_2~.
\end{split}
\end{equation}
It is worth pointing out that the general conditions for the existence of supersymmetric $AdS_7$ solutions of type II supergravity were first derived in \cite{Apruzzi:2013yva}.\footnote{Similar solutions were studied also in earlier work \cite{Blaback:2011pn} where the authors write down a general $AdS_7$ Ansatz in massive type IIA supergravity and find non-supersymmetric solutions of this type.} In Appendix \ref{AFRT} we show that the background in \eqref{metricfinal} together with the differential equations in \eqref{AdsODE} provide a solution to the system of differential equations derived in \cite{Apruzzi:2013yva}. 
\begin{figure}[h!]
\begin{center}
\raisebox{0.65\height}{\includegraphics[width=0.4\textwidth]{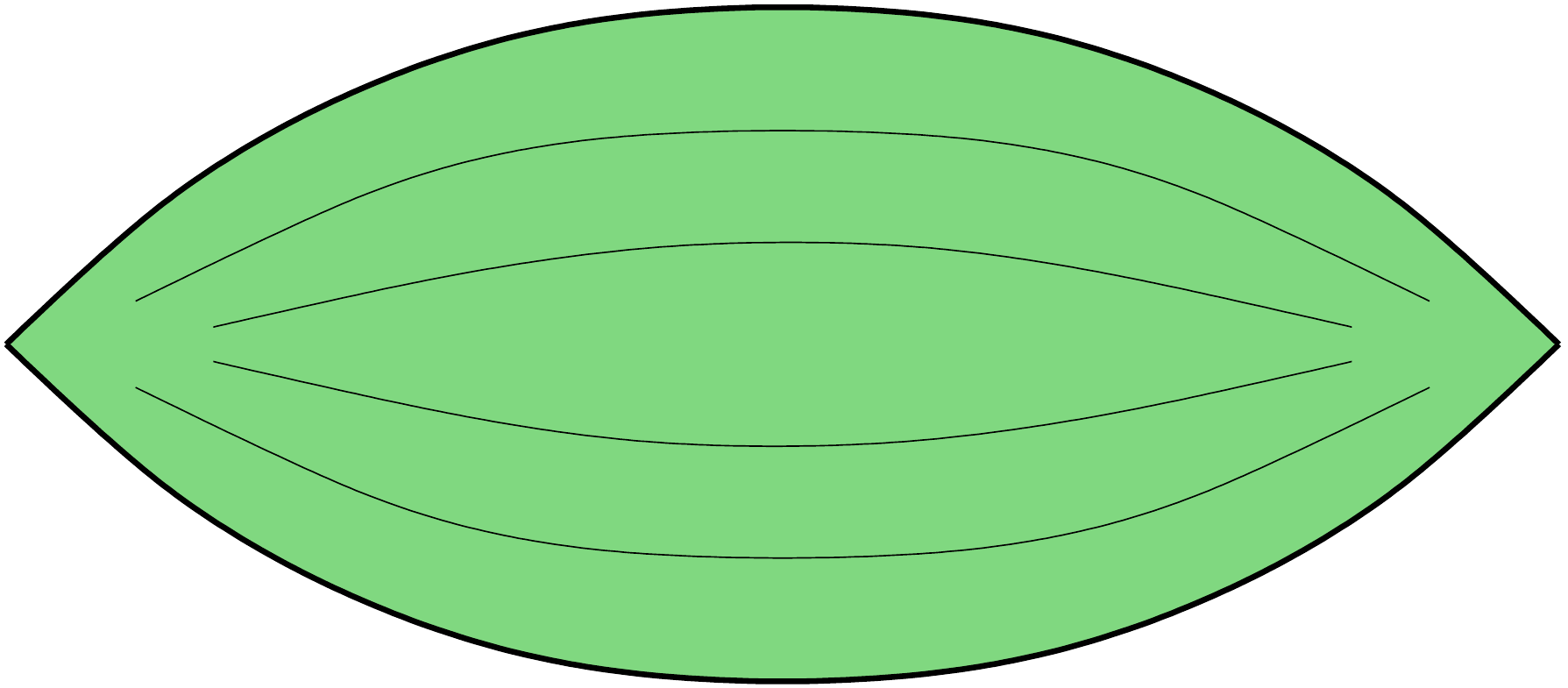}}\qquad\includegraphics[width=0.5\textwidth]{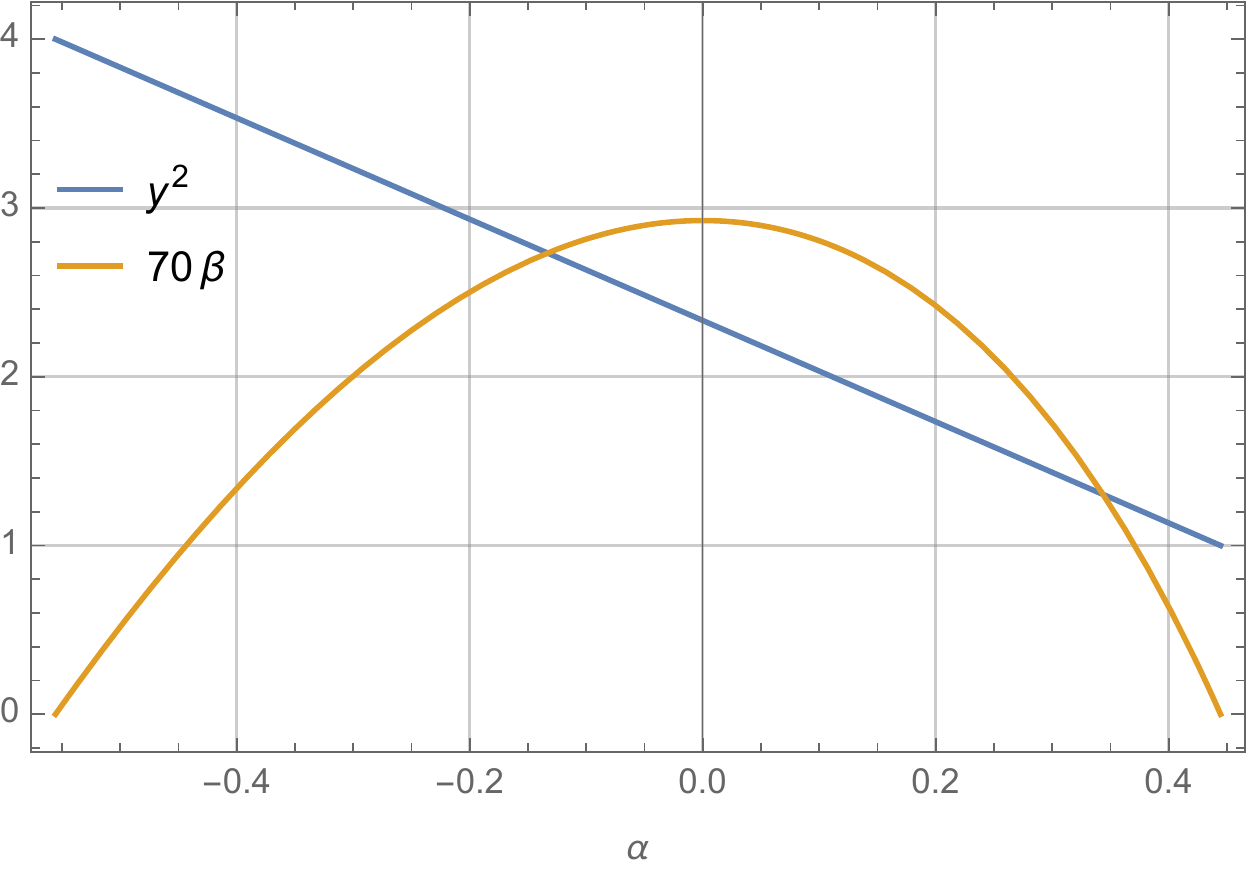}
\caption{\label{fig:twoD6}  The solution (\ref{massivey}-\ref{cubic}) with $M g_s=-3$, $c_1=7/3$ and $c_2 = -6$. The coordinate range for $\alpha$ is $[-5/9,4/9]$. Notice that $y^2$ is a decreasing function of $\alpha$ because of the negative mass and that it takes non-zero values at both poles indicating the presence of D6-branes at both poles. } 
\end{center}
\end{figure}

We end this section with a discussion on the solutions of the differential equations in \eqref{AdsODE} and their interpretation in terms of branes.\footnote{Analytic $AdS_7$ solutions were constructed also in \cite{Rota:2015aoa,Apruzzi:2015zna} and further analyzed in \cite{Cremonesi:2015bld}. Further numerical analysis of such solutions can be found in \cite{Apruzzi:2013yva}.} In the absence of D8-brane charge we have $M=0$ and the solution to \eqref{AdsODE} is 
\be\label{masslessAds}
\beta = \f{1}{4y}(c_2^2-\alpha^2)~,\qquad y = \f{g^2 N_6 g_s}{2\pi}\equiv \sqrt{c_1}~,
\ee
where $c_1$ and $c_2$ are integration constants. One can show that this is the dimensional reduction of the well-known $AdS_7\times S^4/{\mathbb{ Z}}_{N_6}$ supersymmetric background of eleven-dimensional supergravity to type IIA supergravity.\footnote{The ${\mathbb{ Z}}_{N_6}$ orbifold acts on $S^4$ in a way that preserves 16 of the 32 supercharges of $AdS_7\times S^4$.} The coordinate range for $\alpha$ is determined by positivity of $y\beta$ which shows that $-c_2\le\alpha\le c_2$. The solution possesses a D6-brane singularity at $\alpha=\pm c_2$ and the D6 charge at these points is determined by the value of $y$ there as shown in \eqref{masslessAds}. The NS5 brane charge is controlled by the parameters $c_2$ and $g$ which is related to the $AdS_7$ scale via $g=2/L$.

\begin{figure}[h!]
\begin{center}
\raisebox{0.22\height}{\includegraphics[width=0.4\textwidth]{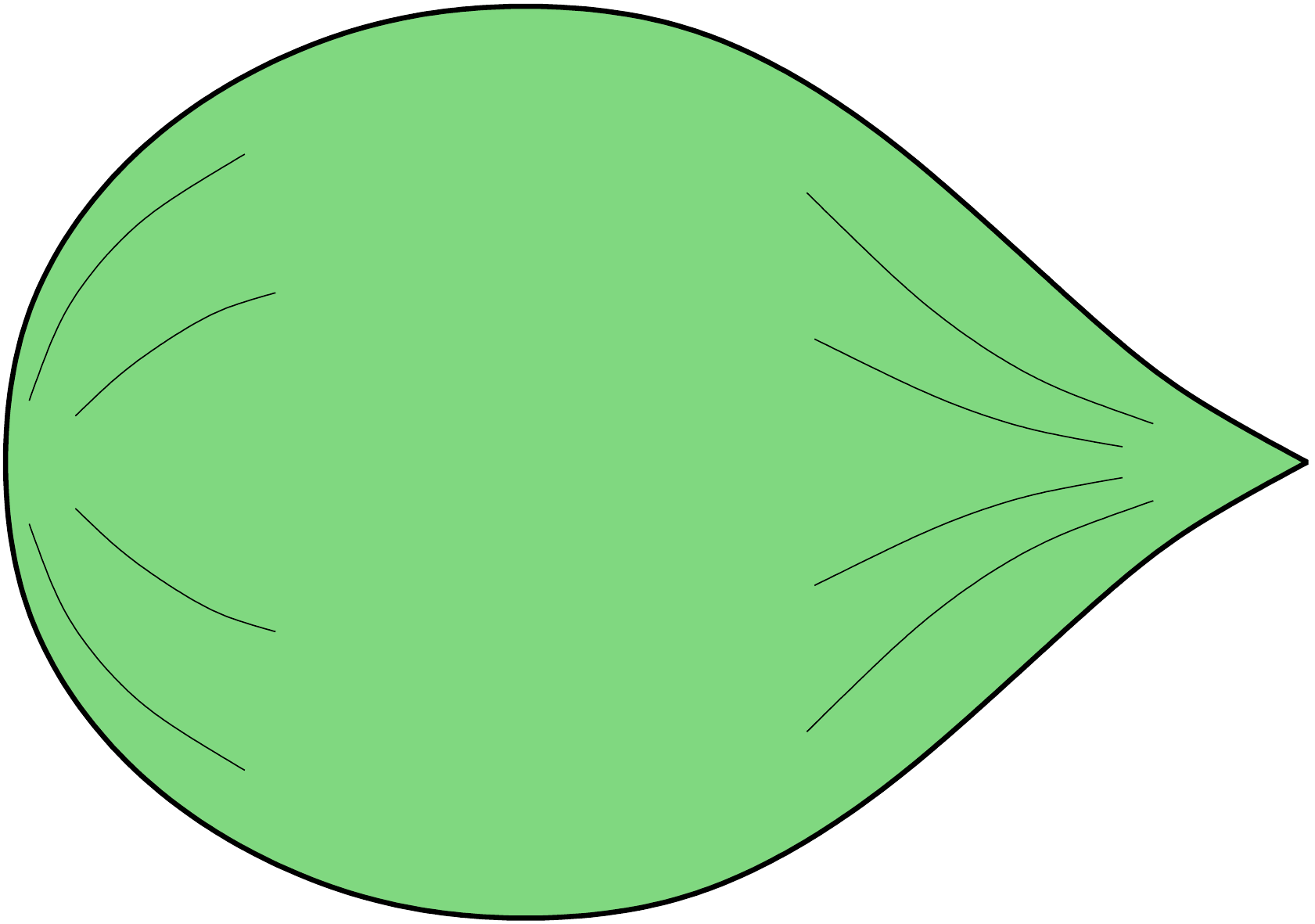}}\qquad\includegraphics[width=0.5\textwidth]{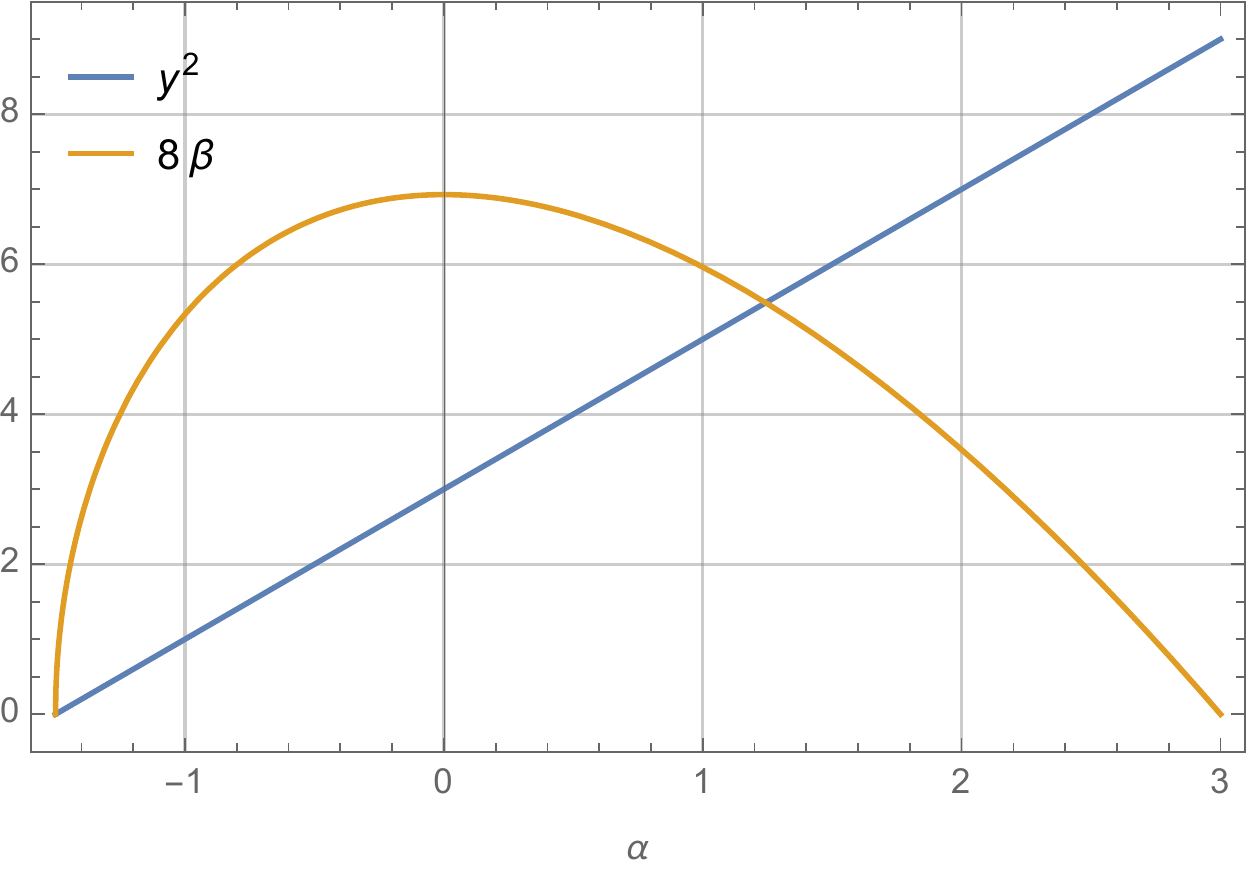}
\caption{\label{fig:oneD6} An example of the solution of (\ref{massivey}-\ref{cubic}) with $M g_s=2$, $c_1=3$ and $c_2 = 0$. The geometry has a stack of D6 branes at one pole, $\alpha=3$, but is regular at the other pole, $\alpha=-3/2$. The function $y(\alpha)$ has a non-zero value at $\alpha=3$ but vanishes at $\alpha=-3/2$ indicating that only one of the poles has D6 branes.} 
\end{center}
\end{figure}
\begin{figure}[h]
\begin{center}
\raisebox{0.25\height}{\includegraphics[width=0.4\textwidth]{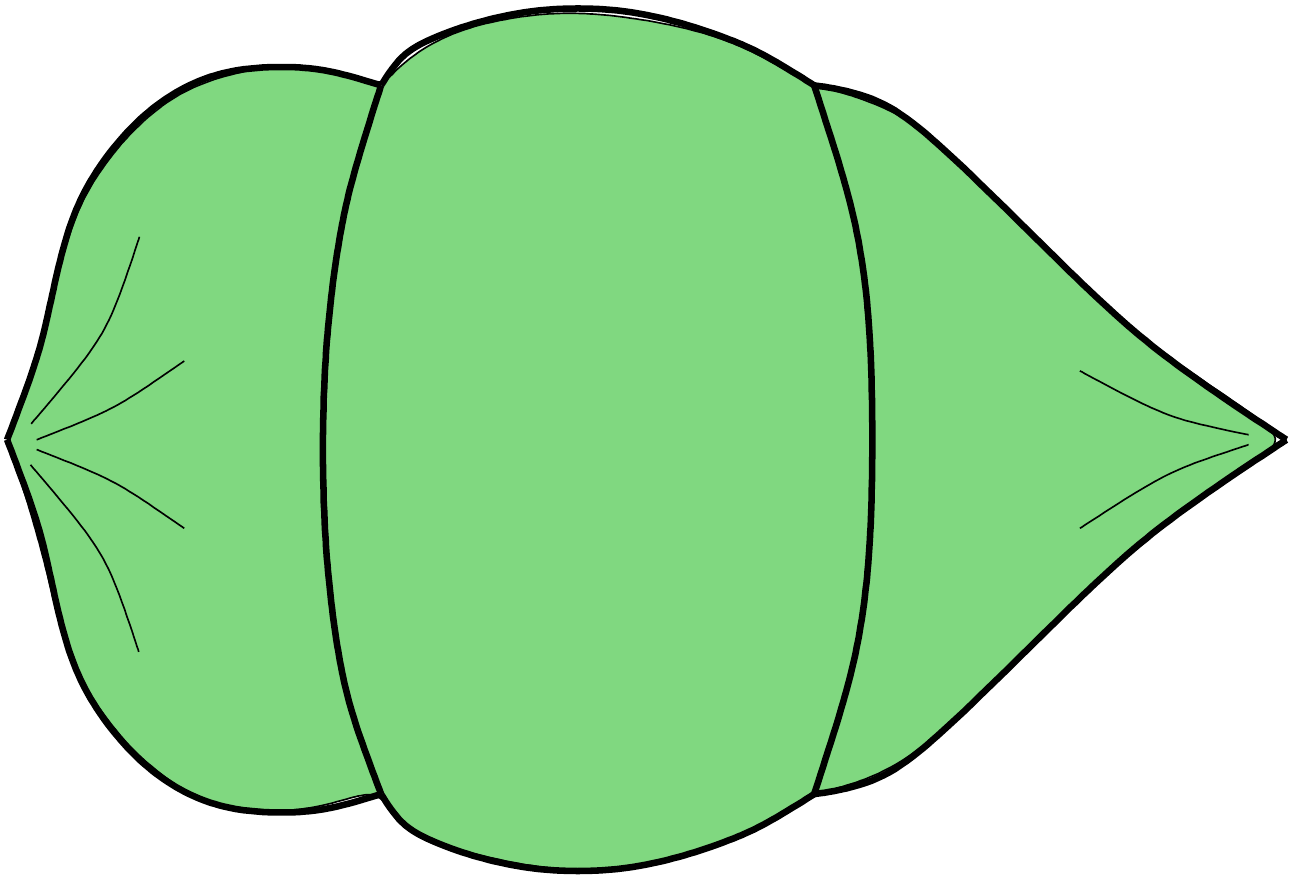}}\qquad\includegraphics[width=0.5\textwidth]{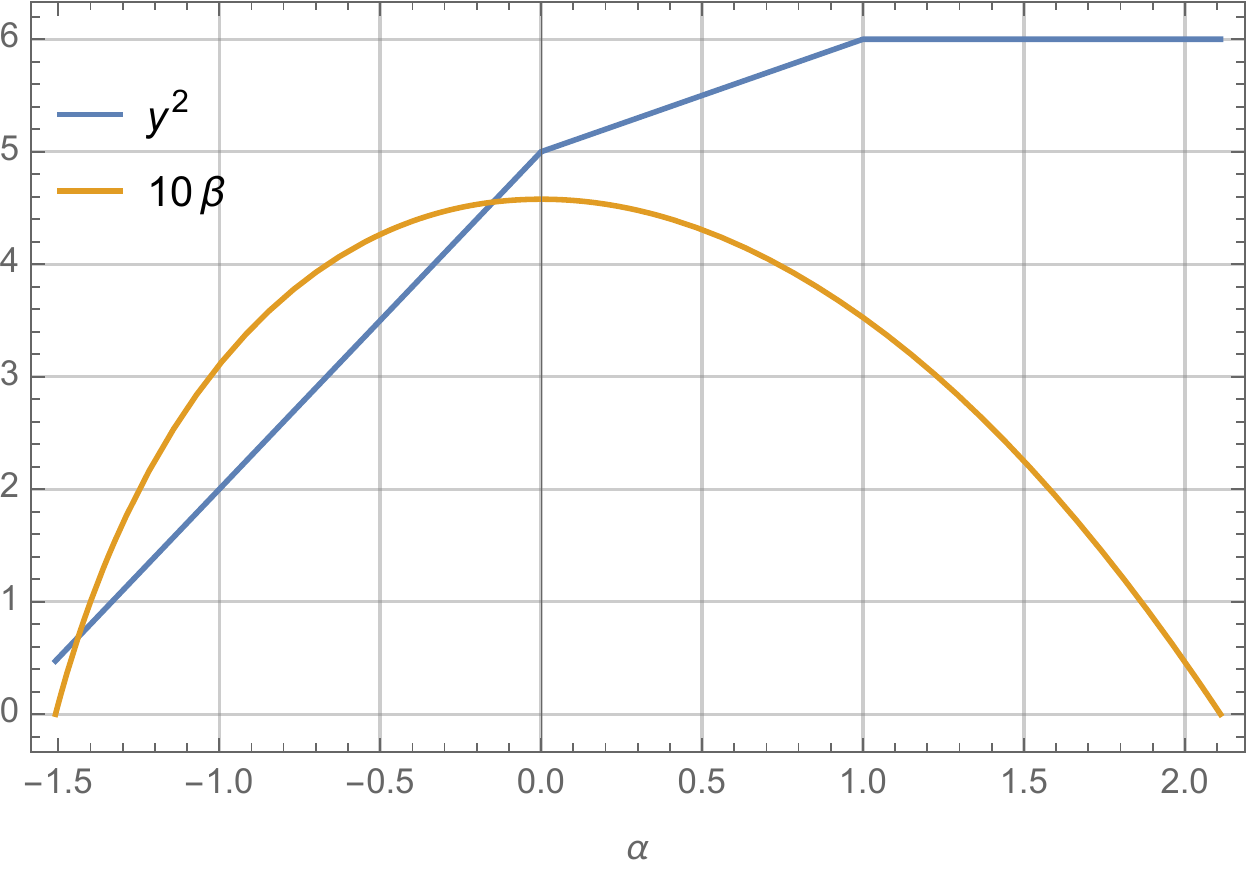}
\caption{\label{fig:2D8}A solution of \eqref{AdsODE} with two D8 brane singularities. The mass parameters are $M^{(1)}g_s=3$, $M^{(2)}g_s=1$ and $M^{(3)}g_s=0$ and determine the slope of the linear function $y^2$. The other integration constants are $c_1^{(1)}=c_1^{(2)}=c_1^{(3)}-1=5$, $c_2^{(1)}=-10$ and $\alpha_+^{(1)}=\alpha_+^{(2)}-1=0$. The remaining constants can be obtained by the continuity of $\alpha$, $\beta$ and $y$. The coordinate $\alpha$ ranges from $\alpha_-^{(1)}\approx -1.51$ to $\alpha^{(3)}_+ \approx 2.11$. The reason, only approximate values are given is that these are obtained by setting $\beta(\alpha)=0$ and are therefore solutions to cubic and quadratic equations respectively.} 
\end{center}
\end{figure}
In general, for non-vanishing Romans mass, $M\ne 0$, $y^2$ is a linear function of $\alpha$,
\be\label{massivey}
y^2 = M g_s \alpha +  c_1~,
\ee
where $c_1$ is an integration constant. One can then solve the second equation in \eqref{AdsODE} in terms of a cubic polynomial in $y$:
\be\label{cubic}
\beta = \f{P(y)}{3(M g_s)^2}~,\quad\text{where}\quad P(y) = -y^3 + 3c_1 y + c_2~,
\ee
where $c_2$ is another integration constant. The same principles hold here as for the massless solution. The coordinate range of $\alpha$ is determined by the positivity of the function $y(\alpha)\beta(\alpha)$, i.e. the positivity of the polynomial $yP(y)$. When $P(y)$ has positive discriminant, $\Delta \equiv 27(4 c_1^3 - c_2^2)$, it has two non--negative roots, and $y$ takes values between these roots. This solution also possesses D6-brane singularities at the ends of the coordinate range and the D6 charge is determined by the value of $y$ at the singularity (See Figure \ref{fig:twoD6} for an example). In the special case when $c_2=0$, one of the roots of $P(y)$ is at $y=0$. In this case the D6 charge there vanishes and the metric is regular (See Figure \ref{fig:oneD6}). If the discriminant $\Delta$ is negative the polynomial $P(y)$ has only one real root and the coordinate range is between $y=0$ and the root of $P(y)$, where one again finds a localized D6-brane singularity. This guarantees that the metric has the correct signature and the dilaton is real.  This solution is once again singular at $y=0$, however in this case the singularity is an O6-plane. Finally, if one has $\Delta=0$, then one finds $c_2^2=4c_1^{3}$ and $c_1>0$.\footnote{The case $c_1=c_2=0$ leads to an unphysical solution.} In this case $P(y)$ has a double root at $y=-\sqrt{c_1}$ and a single root at $y=2\sqrt{c_1}$. Imposing that the dilaton is real and the correct signature of the metric leads to the range $y\in [0,2\sqrt{c_1}]$. At $y=0$ one has an O6-plane singularity and the singularity at $y=2\sqrt{c_1}$ corresponds to a localized D6-brane.

Finally, local solutions with different mass parameters $M$ can be patched together after imposing continuity of $\alpha$, $\beta$ and $y$. The patching surfaces where the value of $M$ changes discontinuously are D8-brane singularities \cite{Imamura:2001cr,Apruzzi:2013yva}. In fact these D8-branes are dielectric, they carry D6 charge and can be understood through the Myers effect \cite{Myers:1999ps} as polarized D6-branes as a result of the $H$-flux in the background \cite{Junghans:2014wda}. The supergravity solution is built by specifying the values of the mass parameter $M$, the integration constants $c_1,c_2$ and the coordinate endpoints $\alpha_-$ and $\alpha_+$ for each region of constant mass parameter. We label these constants in each region by the superscript $(i)$ where $i$ runs over the number of regions $n$. An overall shift in the coordinate $\alpha$ together with the constants $c_1^{(i)}$ enables us to shift the coordinate range and hence $\alpha_-^{(1)}$ can be chosen to take any convenient value. The other parameters $\alpha_\pm^{(i)}$ are related by the continuity constraint $\alpha_+^{(i)} = \alpha_-^{(i-1)}$. The total number of constants to be specified a priori is $4n$. Imposing continuity of $y$ and $\beta$ leads to $2n-2$ constraint equations which in turn reduces the number of free parameters in the solution to $2n+2$. The physical quantities determined by these constants are the $n$ mass parameters $M^{(i)}$, the $n$ dielectric D6 charges embedded in the D8-branes and the two D6 charges at end points, $\alpha_{-}^{(1)}$ and $\alpha_{+}^{(n)}$, of the $\alpha$ interval. An example of a solution with two D8-branes in shown in Figure \ref{fig:2D8}.

\section{Holographic RG flows}
\label{sec:7dflows}

After having shown how to construct the supergravity $AdS_7$ solutions dual to the six-dimensional SCFTs discussed in Section \ref{sec:6dSCFTs} we are now ready to study a class of deformations of these theories which have a universal supergravity description.  These deformations are described by a particular vacuum expectation value (vev) in the field theory that parametrizes a direction in the tensor branch of the vacuum moduli space. Constructing the gravitational dual description of this deformation directly in type IIA supergravity is in general a hard task. Here we sidestep this difficulty by exploiting a seven-dimensional effective supergravity description. It was shown in \cite{Passias:2015gya} (see also \cite{Apruzzi:2016rny}) that supersymmetric vacua of type IIA supergravity of the kind discussed in Section \ref{subsec:AdS7} admit a consistent truncation to a simple seven-dimensional theory known as minimal seven-dimensional supergravity. It is important to emphasize that the details of the particular $AdS_7$ vacuum of IIA supergravity are not visible in the seven-dimensional theory and are encoded in the way one uplifts seven-dimensional solutions to ten dimensions.

As we show below the universal tensor branch deformation of the SCFT is described by a simple supersymmetric domain wall solution of the seven-dimensional supergravity. Similar domain wall ``Coulomb branch'' flow solutions and their holographic interpretation were studied in \cite{Kraus:1998hv,Freedman:1999gk,Cvetic:1999xx}. In particular in \cite{Cvetic:1999xx} (see also \cite{Bakas:1999fa}) the authors focused on domain wall solutions of the maximal seven-dimensional $SO(5)$ gauged supergravity. Thus the solutions they studied are holographically dual to deformations of the interacting $(2,0)$ SCFT living on the worldvolume of coincident M5-branes. The solution we describe below can be obtained as a limit of the solutions of \cite{Cvetic:1999xx} since the seven-dimensional minimal gauged supergravity is a consistent truncation of the maximal theory studied in \cite{Cvetic:1999xx}.

The bosonic sector of minimal supergravity in seven dimensions consists of the metric, a real scalar $\lambda$, a 3-form ${\cal A}_3$ with field strength $\mathcal{F}_4$ and three gauge fields ${\cal A}^I_1$, with field strengths $\mathcal{F}_2^I$, transforming in the adjoint of $SU(2)$ . The bosonic action was originally derived in \cite{Townsend:1983kk}. Here we use the conventions\footnote{We have fixed $h=\frac{g}{2\sqrt{2}}$ in the notation of \cite{Passias:2015gya}.} of \cite{Passias:2015gya} 
\begin{equation}\label{eq:7daction}
\begin{split}
S =& \int d^7x \sqrt{-g_7}\left\{R_7-\f12 |\dd \lambda|^2 - V(\lambda)-\f12 X^4|{\cal F}_4|^2 - \f12 X^{-2} \text{Tr}(|{\cal F}_2|^2)\right\}\\
& +\f12\int \left[ \text{Tr}({\cal F}_2\w {\cal F}_2)-g {\cal F}_4\right]\w{\cal A}_3~,
\end{split}
\end{equation}
where
\be\label{VXdef}
V(\lambda) \equiv -\f12g^2 \left(8X^2 + 8X^{-3} - X^{-8}\right)\quad\text{and}\quad X \equiv \e^{\f{\lambda}{\sqrt{10}}}~.
\ee
The potential can be written in terms of a superpotential as
\begin{equation}\label{eq:7dpotential}
V = \frac{1}{2}\left(\partial_{\lambda} W\right)^2 -\frac{3}{10} W^2\;,
\end{equation}
where we have defined the superpotential
\begin{equation}
W \equiv  g\left(4\;\e^{\frac{\lambda}{\sqrt{10}}}+\e^{-\frac{4\lambda}{\sqrt{10}}}\right)\;.
\end{equation}
There are two $AdS_7$ vacua of this theory which can be found by solving the equation $\partial_{\lambda}V=0$. If an $AdS_7$ vacuum in addition obeys the relation $\partial_{\lambda}W=0$ it preserves some supersymmetry. The vacuum at 
\begin{equation}\label{eq:susyAdS7}
\lambda=0\;, \qquad\qquad V(0) = -\frac{15}{2}g^2\;,
\end{equation}
is supersymmetric and thus perturbatively stable. The dimensionless mass of the scalar $\lambda$ around this vacuum is $m^2L^2= -8$ where $L=2/g$ is the $AdS_7$ scale. This mass is above the BF bound $m_{BF}^2L^2 = -9$ as required for perturbative stability. Using the standard holographic relation
\begin{equation}
\Delta (\Delta-6) = m^2L^2\;,
\end{equation}
we can conclude that the operator $\mathcal{O}_{\lambda}$ dual to the scalar $\lambda$ in the supersymmetric 6d SCFT has dimension $\Delta=4$. In fact $\mathcal{O}_{\lambda}$ is the same as the scalar operator, called $\mathcal{O}$ in Section \ref{sec:6dSCFTs}, in the energy-momentum tensor multiplet and it exists in every $(1,0)$ SCFT.  The $SU(2)$ gauge symmetry is preserved in this vacuum and it is mapped, via the standard holographic dictionary, to the $SU(2)$ R-symmetry in the dual SCFT. 

The other $AdS_7$ vacuum of the minimal gauged supergravity is at
\begin{equation}
\lambda=\lambda_*=-\frac{2}{\sqrt{10}} \log(2)\;, \qquad\qquad V(\lambda_*) =  -5 \times 2^{3/5}g^2\;.
\end{equation}
The $AdS_7$ scale is $L_{*} =  2^{1/5} \times 3^{1/2}/g$ and one finds that the mass of $\lambda$ is $m_{*}^2L_{*}^2= 12$. This means the the scalar operator in the dual CFT is irrelevant with conformal dimension $\Delta_{*} = 3+\sqrt{21} \approx 7.58$. This vacuum does not preserve any supersymmetry and is perturbatively stable within the minimal seven-dimensional supergravity as well as in the supergravity theory discussed in \cite{Apruzzi:2016rny}. It is however a perturbatively unstable vacuum of the maximal seven-dimensional $SO(5)$ gauged supergravity as shown in \cite{Pernici:1984zw}. This vacuum will not play any further role in our discussion.

The domain wall solution we are interested in can be derived by setting the gauge fields and the 3-form in \eqref{eq:7daction} to zero and using a standard domain wall Ansatz for the metric and scalar field
\begin{equation}\label{DWansatz}
\dd s^2_7 = \dd\eta^2 +\e^{2\mathcal{A}(\eta)}\dd s^2_6\;, \qquad \lambda(\eta)\;.
\end{equation}

We would like to emphasize an important point for our further analysis. Any solution of the minimal seven-dimensional supergravity of the form \eqref{DWansatz} can be uplifted to a solution of massive type IIA supergravity using the results in \cite{Passias:2015gya,Apruzzi:2016rny}. There is some freedom in the way this uplift is performed which is encoded in the cubic polynomial $P(y)$ introduced in eq. \eqref{cubic}. As explained there, $P(y)$ is only piecewise cubic and the singularities of $P(y)$ determine the location of D8 branes where the mass parameter changes value. Here we will stick to a fixed mass, $M$, and will choose $P(y)$ to be a cubic polynomial. The extension to include D8 branes is straight forward. Using the results in \cite{Passias:2015gya} adapted to our notation we find that the full type IIA supergravity background corresponding to a seven-dimensional solution of the type \eqref{DWansatz} is
\begin{equation}\label{metricnew}
\begin{split}
\dd s^2 &= \sqrt{\f{\beta}{X y}}\left\{\dd s^2_{7} + \f{X^3}{g^2\beta y}\left(
\f{\dd \alpha^2}{4} + \f{(\beta y)^2}{\alpha^2+4X^5 y\beta}~\dd\Omega^2_2\right)~\right\},\\
e^{4\phi} &= \f{16 g^4 g_s^4 \beta^3 X^4}{y^3(\alpha^2+4X^5 y\beta)^2}~,\\
F_2 &= \f{1}{2g^2g_s}\left(y+ \f{M g_s \beta\alpha}{\alpha^2 + 4X^5y\beta} \right)~\vol_2~,\\
H &= \f{\beta}{2g^2y(\alpha^2 + 4 X^5 y \beta)} [(2X^5+1)y\dd\alpha\\ 
&\qquad\qquad\qquad \qquad~-2\alpha\f{(2-X^5)M g_s \beta\dd\alpha+ 2\dd\left(y^2\beta(X^5-1)\right)}{\alpha^2 + 4X^5 y \beta} ]\w\vol_2\;,
\end{split}
\end{equation}
where $\dd s_7^2$ is the metric in \eqref{DWansatz} and $X$ is the scalar field as defined in \eqref{VXdef}. The functions $y$ and $\beta$ satisfy the same equations \eqref{AdsODE} as for the undeformed $AdS_7$ backgrounds. For a fixed mass $M$ they are given by \eqref{massivey} and \eqref{cubic}.

To find supersymmetric domain wall solutions of the form \eqref{DWansatz} we plug this Ansatz in the supersymmetry variations of the seven-dimensional theory and find that any background of this type should obey the following differential equations:
\begin{equation}\label{eq:7dBPS}
\begin{split}
\frac{\dd \lambda}{\dd \eta} &= -\frac{\partial W}{\partial \lambda}\;, \\
\frac{\dd \mathcal{A}}{\dd \eta} &= \frac{1}{10} W\;.
\end{split}
\end{equation}
To solve this system of equations we find it convenient to perform the following change of variables:
\begin{equation}
\dd \eta = \frac{\rho\dd \rho}{g((\rho^2-\ell_1^2)^4(\rho^2-\ell_5^2))^{1/5}}\;.
\end{equation}
With this at hand one can then solve the system of equations in \eqref{eq:7dBPS} analytically. We will omit the derivation here and only quote the result using notation which fits in the general framework studied in \cite{Cvetic:1999xx}. The non-trivial fields are
\begin{equation}\label{7dsolution}
\begin{split}
\dd s_7^2 =&\f{1}{(g\rho)^2 (H_1^4H_5)^{2/5}} \dd \rho^2 +(g \rho) (H_1^4H_5)^{1/10}  \dd s_6^2~,\\
X(\rho)^5 =& \f{H_5}{H_1}~, 
\end{split}
\end{equation}
where
\be\label{7dharmonics}
H_1(\rho) = 1 - \f{\ell_1^2}{\rho^2}~,\quad H_5(\rho) = 1 - \f{\ell_5^2}{\rho^2}~.
\ee
Note that we have used the notation of \cite{Cvetic:1999xx} which is adapted to treating similar domain walls in the maximal seven-dimensional $SO(5)$ gauged supergravity. In particular we have set $\ell_1=\ell_2=\ell_3=\ell_4$ in the notation of \cite{Cvetic:1999xx} thereby making four of the five scalars considered there equal.\footnote{We have taken the integration constants in \eqref{7dharmonics} to be negative to make the singularity at the end of the flow apparent.} We choose to present the seven-dimensional domain wall solution in this language in order to make contact with the uplifted supergravity solution discussed in Section \ref{subsec:interpretM5s} below. 

In the limit $\rho\to\infty$ the metric reduces to AdS$_7$ in the vacuum \eqref{eq:susyAdS7}, it is convenient to change coordinates in this limit
\be
g\rho = \e^{g\eta}~,
\ee
such that the metric takes the form
\be
\dd s_7^2 = \dd \eta^2 + \e^{2 \eta/L}\dd s_6^2~,\quad\text{where}\quad L = \f{2}{g}~.
\ee
The canonically normalized scalar field in this limit has the expansion
\be\label{eq:lambdaUV}
\lambda = g^2\f{2}{\sqrt{10}}(\ell_1^2-\ell_5^2)\e^{-4\eta/L} + g^4\f{1}{\sqrt{10}}(\ell_1^4-\ell_5^4)\e^{-8\eta/L}+\cdots~.
\ee
Since the operator dual to $\lambda$ is of dimension $4$, the coefficient of $\e^{-4\eta/L}$ is proportional to the vev, $v$, of the dual operator where 
\be\label{theVEV}
v \equiv g^2\f{2}{\sqrt{10}}(\ell_1^2-\ell_5^2)~.
\ee
The source is given by the coefficient of $\e^{-2\eta/L}$ in the UV expansion of the scalar \eqref{eq:lambdaUV}, and hence vanishes. The fact that the source term in \eqref{eq:lambdaUV} vanishes is in harmony with the results of \cite{Cordova:2016xhm} where it was shown that the only supersymmetric relevant deformations of six-dimensional SCFT are given by vevs.

It is clear that at values of $\rho$ where either $H_1$ or $H_5$ vanishes, the metric is singular. The range of the coordinate $\rho$ is therefore set by the larger of the two integration constants $\ell_1^2$ and $\ell_5^2$. Without loss of generality we can choose $\ell_1$ and $\ell_5$ to be positive and thus we find the coordinate range
\be
\max\{\ell_1,\ell_5\}\le \rho\le  \infty~.
\ee
The nature of the curvature singularity encountered at the minimum value of $\rho$ depends on which of the two integration constants, $\ell_1$ or $\ell_5$, is greater. We explore both possibilities below. When $\ell_1< \ell_5$ the metric locally takes the form as $ \rho\to \ell_5$
\be\label{eq:sing1}
\dd s^2_7\approx\dd \zeta^2 + \sqrt{-v}\left(10~g \zeta\right)^{1/8}\dd s_6^2~,
\ee
where we changed coordinates as follows
\be
\rho-\ell_5 = -\f{v}{\ell_5}\left(\f{8}{5g}\right)^{3/4} \zeta^{5/4}~.
\ee
When $\ell_1>\ell_5$ the metric locally takes the form
\be\label{eq:sing2}
\dd s^2_7 \approx \dd \zeta^2 +\sqrt{v}\left(\f25\right)^{7/4}g^2 \zeta^2 \dd s_6^2~,
\ee
where we have defined
\be
\rho - \ell_1 = \f{\sqrt{10}~v}{\ell_1}\left(2g\right)^3 \left(\f{\zeta}{5}\right)^{5}~.
\ee
Finally when $\ell_1=\ell_5$ the solution trivializes. The scalar is constant, $X=1$, and the metric is that of the supersymmetric $AdS_7$ vacuum in \eqref{eq:susyAdS7}.

The metrics in \eqref{eq:sing1} and \eqref{eq:sing2} have a curvature singularity and are therefore hard to interpret in the realm of classical supergravity. Fortunately holography and string theory have offered insights into this type of singularities. In particular there are two well-known criteria for deciding which curvature singularities arising in similar holographic domain walls are acceptable \cite{Gubser:2000nd,Maldacena:2000mw}. The criterion in \cite{Gubser:2000nd} states that a singularity is acceptable only if the scalar potential in \eqref{eq:7dpotential} is bounded from above. It is easy to verify that this the case only when $\ell_1\ge \ell_5$. The Maldacena-Nu{\~n}ez criterion states that for acceptable singularities in string theory the $g_{tt}$ component of the ten-dimensional Einstein frame metric should be bounded above as the singularity is approached. The results in Section \ref{subsec:interpretM5s} and Section \ref{subsec:interpretD6NS5s} show that applying this criterion again leads to the condition $\ell_1\ge \ell_5$ for a physically acceptable singularity. From now on we will therefore take $\ell_1\ge \ell_5$ which, using \eqref{theVEV}, is equivalent to
\be
v\ge 0~.
\ee
This result is in harmony with the field theory discussion below \eqref{eq:6dLag}. The parameter $v$ is dual to the vev of a scalar operator that parametrizes a particular direction on the tensor branch. When $v\neq 0$ one may think of this vev as the effective gauge coupling on some locus of the tensor branch, $v \sim 1/g_{YM}^2$. The constraint $v\ge 0$ therefore agrees with this intuition since it implies that the effective couplig $g_{YM}^2$ is positive.

We would  like to end this section with a technical comment that will play a role in the subsequent discussion. As explained in \cite{Cvetic:1999xx}, the smaller of the integration constants $\ell_{1,5}$ can be shifted to zero by a redefinition of the coordinate $\rho$. This amounts to shifting all the integration constants by the smallest one. We will make use of this result to eliminate the constant $\ell_5$. The result of this choice is that the seven-dimensional metric and scalar take the same form as before \eqref{7dsolution}, but the functions $H_1$ and $H_5$ are now
\be\label{7dharmonicsafterrescaling}
H_1(\rho) = 1- \f{\ell_1^2-\ell_5^2}{\rho^2} = 1- \f{\sqrt{10}~v}{2(g\rho)^2}~,\qquad H_5(\rho)=1~,
\ee
where we have also made use of \eqref{theVEV}.

\subsection{Uplift to eleven dimensions}
\label{subsec:interpretM5s}

Before interpreting the domain wall in terms of intersecting NS5- and D6-branes in massive type IIA, we first review how the solution can be uplifted to eleven dimensions and interpreted as a distribution of M5-branes. The eleven dimensional metric takes the standard M5 brane form \cite{Cvetic:1999xx} (see also \cite{Bakas:1999fa}) 
\be
\dd s_{11}^2 = h^{-1/3} \dd s_6^2 + h^{2/3}\dd s_5^2~,
\ee
where
\bea
h^{-1} &=& (g \rho)^3 (H_1H_2H_3H_4H_5)^{1/2}\sum_{i=1}^{5}H_i^{-1}\mu_i^2~,\\
\dd s_5^2 &=&  \sum_{i=1}^{5}\left(H_i^{-1}\mu_i^2 \dd \rho^2 + \rho^2 H_i\dd\mu_i^2\right)~.
\eea
For the domain wall \eqref{7dsolution} the harmonic functions are $H_5=1$ and  $H_1=H_2=H_3=H_4$ is given in \eqref{7dharmonicsafterrescaling}. The coordinates $\mu_i$ parametrize a four-sphere and satisfy $\sum_{i=1}^{5}\mu_i\mu_i=1$. By a change of coordinates $y_i \equiv \rho \sqrt{H_i}\mu_i$ the five dimensional metric $\dd s_5^2$ can be made manifestly flat, $\dd s_5^2 = \dd y_i\dd y_i$. It is simple to verify that the function
\be
h = \f{4 g \rho}{(\sqrt{10}~v - 2g^2\rho^2)(\sqrt{10}~v\mu_5^2 - 2 g^2 \rho^2)}~,
\ee
is harmonic, up to isolated singularities, in the five-dimensional space spanned by $(y_1,y_2,y_3,y_4,y_5)$. These singularities determine a distribution of M5-branes
\be
-\triangle_5 h = \sigma_\text{M5}~,
\ee
where $\sigma_\text{M5}$ is the charge density of this distribution. The charge density was determined in \cite{Cvetic:1999xx} (using the techniques of \cite{Kraus:1998hv}) to be
\be
\sigma_\text{M5} = \left(\f{2\pi}{g}\right)^2\left(\f{4}{10 v^2}\right)^{1/4}\Theta\left(\sqrt{10}~v - 2 g^2y_5^2\right)\delta^{(4)}(y_1,y_2,y_3,y_4)~.
\ee
Given this charge density the harmonic function $h$ can be written as
\be\label{M5convol}
h = \int E_\text{M5} (\vec{y}-\vec{y}') \sigma_\text{M5} (\vec{y}') d\vec{y}'~,
\ee
where $E_\text{M5} \equiv (8\pi^2 (y_iy_i)^{3/2})^{-1}$ is the fundamental solution to the Laplace equation in the flat five-dimensional space spanned by $y_i$.

The eleven-dimensional domain wall solution presented above should be the gravitational dual to a particular direction in the tensor branch of the six-dimensional $(2,0)$  superconformal theory of type $A_{N_5}$ that lives on the worldvolume of $N_5$ M5 branes. It should also capture an analogous locus on the tensor branch of the $(1,0)$ cousins of this $(2,0)$ SCFTs which are obtained by placing coincident M5 branes at ADE singularities. See \cite{DelZotto:2014hpa} for a recent discussion of the tensor branch of these $(1,0)$ SCFTs. It will be very interesting to make the correspondence between holography and field theory on this branch of the moduli space more precise.

\subsection{Uplift to type IIA supergravity}
\label{subsec:interpretD6NS5s}

The seven-dimensional domain wall flow solution in \eqref{7dsolution} can also be uplifted to massive type IIA supergravity via the uplift formulas presented in \eqref{metricnew} which were derived in \cite{Passias:2015gya} (see also \cite{Apruzzi:2016rny}). The uplifted solutions can be cast into the ``intersecting brane'' form \eqref{ansatz1}-\eqref{ansatz4} for which the metric takes the form
\be
\dd s^2 = S^{-1/2} \dd s_6^2 + K\left[ S^{-1/2} \dd z^2 + S^{1/2} (\dd r^2 + r^2 \dd \Omega_2^2)\right]~.
\ee
The only technical task is to determine how the coordinates $r$ and $z$ which are natural in \eqref{ansatz1}-\eqref{ansatz4} get mapped to the coordinates $\rho$ and $\alpha$ in the uplift formulas \eqref{metricnew}.

We start by analyzing this ten-dimensional uplifted solution for vanishing Romans mass, $M=0$. Then we have  
\be\label{eqSM0}
S(r) = \f{N_6 g_s}{4\pi r}~,
\ee
where the parameter $N_6$ appears as a free constant compatible with the uplift formulas, i.e. it is not determined in terms of any quantity in the seven-dimensional flow solution. The function $K$ is
\be\label{masslessKfunc}
K = \f{16 g^3\rho}{(\sqrt{10}~ v - 2g^2\rho^2)(\sqrt{10}~v \alpha^2-2c_2^2 g^2\rho^2)}~,
\ee
where $\alpha\in[-c_2,c_2]$ was defined in \eqref{alphacoordinate} and $\rho$ is the seven-dimensional radial coordinate as in \eqref{7dsolution}. With this at hand we find the following relation between the coordinates $(r,z)$ and $(\rho,\alpha)$
\be
r=(\sqrt{10}~ v - 2g^2\rho^2)\f{\pi(\alpha^2 -c_2^2)}{8 g^4N_6g_s}~,\qquad z = \f{\rho \alpha}{2g}~.
\ee
As in the case of M5 branes \eqref{M5convol} we can express the harmonic function $K$ in terms of a convolution with the fundamental solution of the last equation in \eqref{imamuraeq}
\be\label{masslessKasconv}
K = \left(\f{N_6 g_s}{4\pi}\right)\int \f{1}{4\pi((z-z')^2 + 4S(r)r^2)^{3/2}}\sigma_\text{NS5}(z')\dd z'~,
\ee
where the charge density is defined as
\be\label{sigma5NS}
\sigma_\text{NS5}(z) \equiv \left(\f{2\pi}{g}\right)^3\left(\f{4}{10v^2}\right)^{1/4}\f{1}{(c_2 N_6 g_s)^2}\Theta\left(\sqrt{10}~v-32 g^3z^2\right)~.
\ee
This in turn implies that equation \eqref{masslessKfunc} provides a solution to the following equation
\be
-\triangle_3 K - S(r)\partial_z^2 K = \sigma_\text{NS5}(z)~.
\ee
This equation is simply the last equation in \eqref{imamuraeq} with a non-trivial source provided by the NS5 charge density $\sigma_\text{NS5}(z)$. It should be noted that this solution is not asymptotically flat in ten dimension. Having found the charge density $\sigma_\text{NS5}$, a full  ten-dimensional asymptotically flat solution will be determined in Section \ref{sec:masslessflatsol}.

We now move to the case in which $M\neq 0$. For the domain wall \eqref{7dsolution} we can express the uplift in terms of the harmonic functions $S$ and $K$ as for the massless case. Once again the D6 ``harmonic function'' takes a simple form
\be\label{eqSMnot0}
S = \f{y}{2 g^2 r}~.
\ee
However, the function $K$ takes a substantially more complicated form
\be\label{eqKMnot0}
K = \f{g^3 (12 M g_s)^2\rho}{(\sqrt{10}~v-2g^2 \rho^2)(\sqrt{10}~vP'(y)^2 - 2g^2 \rho^2(P'(y)^2+ 12yP(y)))}~.
\ee
Here $y$ and $\rho$ are related to  the coordinates $r$ and $z$ through
\be\label{eqrznot0}
r=-(\sqrt{10}~v-2g^2 \rho^2)\f{P(y)}{12(M g_s g)^2}~,\qquad z = -\f{\rho P'(y)}{6 M g_s g}~.
\ee
We should emphasize that the notation we are using here is similar to the one used for the ten-dimensional $AdS_7$ solutions in Section \ref{subsec:AdS7} since the domain wall solutions at hand are deformations of these $AdS_7$ vacua controlled by the parameter $v$. In particular from the three equations in \eqref{thirdeq} only the first one is obeyed by the domain wall with $v\neq 0$ and the other two are broken. This also implies that the function $S$ still has the same form as in equation \eqref{SKyGeq} as is evident from \eqref{eqSM0} and \eqref{eqSMnot0} above.

\section{Asymptotically flat brane intersections for $M=0$}
\label{sec:masslessflatsol}

In this section we focus on massless type IIA supergravity and find explicitly a supergravity solution that completes the intersecting brane solution found in the previous section to a ten-dimensional asymptotically flat background. For the M5 brane solution in eleven--dimensional supergravity, this task is easily accomplished simply by adding a constant to the harmonic function in \eqref{M5convol}
\be
h = 1+  \int E_\text{M5} (\vec{y}-\vec{y}') \sigma_\text{M5} (\vec{y}') d\vec{y}'~.
\ee
In type IIA supergravity in the presence of both NS5- and D6-branes the situation is more complicated. Naively one is inclined to ``add 1'' to both functions $S$ and $K$ in order to recover the elementary D6- and NS5-brane solutions in \eqref{D6branesol} and \eqref{NS5branesol}. However this procedure does not lead to a solution since the system of BPS equations in \eqref{imamuraeq} are coupled and nonlinear. In the massless limit, $M=0$, of type IIA supergravity the problem however reduces to a linear one which we solve below.

\begin{figure}
\begin{center}
\includegraphics[width=0.85\textwidth]{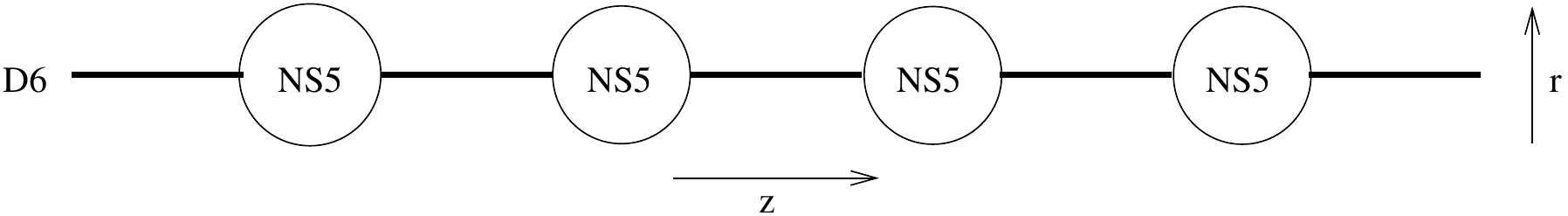}
\caption{\label{intersection}The explicit realisation of intersecting branes in type IIA with $M=0$. A single stack of $N_6$ D6-branes fills the $z$-direction while NS5-branes are scattered over the same direction.}
\end{center}
\end{figure}

Inspired by the solution obtained by uplift in \eqref{eqSM0}-\eqref{sigma5NS} we study the intersection of D6 and NS5 branes for which the NS5 are localized at $r=0$ but spread along the $z$-direction. The stack of D6-branes is kept localized at $r=0$ (see figure \ref{intersection}). A solution of this type but with a single stack of NS5-branes was constructed previously in \cite{Clarkson:2004eq}. We will start by reviewing that solution and then extend it to a distribution of NS5-branes.
Remember that the PDEs in \eqref{imamuraeq} are obtained as a result of the Bianchi identities. Let us set $M=0$ and write these Bianchi identities with explicit brane sources 
\begin{equation}
\begin{split}
\dd F_2 &= -  N_6  \delta(r) ~r^2\dd r\w \vol_2~,\\
\dd H &= -N_5 \delta(r)\delta(z)~r^2\dd z\w\dd r\w \vol_2~.
\end{split}
\end{equation}
The effect of adding explicit brane sources on the right hand side of the Bianchi identities is the following modification of the PDEs in \eqref{imamuraeq}
\bea
\partial_z S &=& 0 ~,\nonumber\\
-\triangle_3 S  &=& g_s N_6\delta(r) ~,\label{imamuraeqwdelta}\\
-\triangle_3 K - \partial_z^2(KS) &=& N_{5} \delta(r)\delta(z)~.\nonumber
\eea
The delta functions serve to fix boundary values of $S$ and $K$ when an explicit solution is written down. The function $S$ is independent of $z$ and is found to be
\be\label{M0Ssol}
S = a_1^2+ \f{N_6g_s}{4\pi r}~,
\ee
where $a_1$ controls part of the asymptotic behavior of the solution. Notice that the uplift of the seven-dimensional domain wall solution lead to the function $S$ in \eqref{eqSM0}, i.e. to $a_1=0$. Here we will explore the more general situation with $a_1\ne0$. 

 The general system of equations in \eqref{imamuraeq} possesses two scaling symmetries. These symmetries act on the fields and coordinates as follows
\be\label{scalings}
\begin{split}
r\to r'=s^2 r~,&\qquad z\to z'=t^2 z~,\\
 x^\mu \to x'^\mu=s^{-1} t x^\mu~,&\qquad g_s\to g_s'=s^{-2}t^4 g_s\\
S\to S'=s^{-4}t^4 S(r,z)~,&\qquad K\to K'=s^{-2}t^{-2} K(r,z)~,
\end{split}
\ee
where $t,s$ are arbitrary real numbers. One of these scaling symmetries can be used to set $a_1 =1$ (as long as $a_1\neq 0$) which we will do from now on. Later on we will be interested in exploring the limit $a_1\to0$ which can be achieved by taking the limit $r\to0$ while keeping $g_s$ finite. With $S$ at hand the function $K$ then satisfies a \emph{linear} PDE
\be\label{masslessKeq}
-\triangle_3 K - S(r)\partial_z^2 K = N_{5} \delta(r)\delta(z)~.
\ee
To solve this equation we proceed by a Fourier transform along the $z$-coordinate:
\be
-\triangle_3 \hat{K} + S(r)\lambda^2 \hat{K} = \f{N_{5} }{\sqrt{2\pi}} \delta(r)~,\label{fourierKeq}
\ee
where we have set
\be
K(r,z) = a_2^2 + \f{1}{\sqrt{2\pi}}\int_{-\infty}^{\infty} \hat{K}(r,\lambda)\e^{i\lambda z}\dd \lambda~,
\ee
where $a_2$ is a constant that will ultimately also control the asymptotic behavior of the solutions. We will use the second scaling symmetry of the system to set $a_2=1$ (again assuming that $a_2\neq 0$). The homogeneous solution to \eqref{fourierKeq} is
\bea\label{Khatgen}
\hat{K} &=& b_1(\lambda)\e^{-|\lambda| r}U\left(1 + \f{N_6 |\lambda| g_s}{8\pi },2,2  |\lambda| r\right)\nonumber\\
&& \qquad\qquad+ \; b_2(\lambda)\e^{-|\lambda| r}{}_1F_1\left(1 + \f{N_6  |\lambda| g_s}{8\pi },2,2 |\lambda| r\right)~,
\eea
where $U$ and ${}_1F_1$ are hypergeometric function. The second term diverges for large $r$ and so we must set $b_2(\lambda)=0$. Once $b_1(\lambda)$ has been determined the full solution is written entirely in terms of $U$ which is defined by
\be
U(a,b,z) \equiv \f{1}{\Gamma(a)} \int_0^\infty \e^{-z\tau}\tau^{a-1}(1+\tau)^{b-a-1}\dd \tau~,\quad a>0~.
\ee
We can determine $b_1(\lambda)$ by integrating  \eqref{fourierKeq} in a ball of radius $\epsilon$ and taking $\epsilon\to0$:
\be
-4\pi \lim_{\epsilon\to0} \int_0^\epsilon r^2  (\triangle_3 \hat{K} - S(r)\lambda^2 \hat{K})\dd r = \f{N_{5}}{\sqrt{2\pi}}  ~.
\ee
Only the first term on the left hand side gives a finite contribution as $\epsilon\to0$ which results in the following equation for $b_1(\lambda)$,
\be
b_1(\lambda) = \lambda^2 \f{N_{5} N_6g_s}{\sqrt{2\pi}~16\pi^2}\Gamma\left(\f{N_6g_s|\lambda|}{8\pi}\right)~,
\ee
where $\Gamma(x)$ is the Euler gamma function. It is easy to see that the limit for which $N_{5}$ vanishes gives the solution for $N_6$ D6-branes given in \eqref{D6branesol}. A slightly more involved limit is $N_6\to 0$ for which $S\to 1$ and 
\be
K\to  1 + \f{N_{5}}{4\pi^2(r^2+z^2)}~, 
\ee  
which is the harmonic function for a collection of NS5 branes in \eqref{NS5branesol}.

We now explore the ``near horizon'' limit $r\to0$ while keeping $g_s$ finite.\footnote{What we refer to as a ``near horizon'' limit can be thought of as a limit in which one zooms in on the NS5 branes in a controlled manner.} The scaling symmetries \eqref{scalings} show that in order to keep $g_s$ finite, $r/z^2$ must also remain finite in this limit. This is in good agreement with the analysis in Appendix \ref{app:AdS7} which shows that for supersymmetric $AdS_7$ solutions the background fields depend nontrivially only on the combination $r/z^2$. In this limit $S(r)$ reduces to
\be
S(r) \to \f{N_6g_s}{4\pi r}~.
\ee
We should expect that $K$ also reduces to its $AdS_7$ form \eqref{SKyGeq}. To evaluate $K$ in the $r\to0$ limit we use a convenient expansion of the $U$-function in terms of the Bessel functions $\mathcal{K}_n$ for large $a$ \cite{TEMME198127}. The first term in this expansion is
\be
U(a,b,z) \approx 2\f{\e^{z/2}}{\Gamma(a)}\left(\f{z}{a}\right)^{(1-b)/2} \mathcal{K}_{1-b}(2\sqrt{az})~.
\ee
Using this in \eqref{Khatgen} we obtain
\be
\hat{K}(r,\lambda) \to |\lambda|\f{N_5 N_6g_s}{\sqrt{2\pi}~16\pi^2} \sqrt{\f{N_6 g_s}{4\pi r}}\mathcal{K}_1\left(\sqrt{\f{N_6 g_s r}{\pi}}|\lambda|\right) ~,
\ee
which has the Fourier transform
\be\label{masslessNS5harmonic}
K(r,z)\to \left(\f{N_6 g_s}{4 \pi}\right)^2\f{N_{5}}{4\pi(z^2 + 4S(r)r^2)^{3/2}}~.
\ee
This solution can now be compared to the pure massless AdS$_7$ solution in \eqref{masslessAds} and indeed we find that \eqref{masslessNS5harmonic} can be written as
\be
4 y \beta = \left(\f{N_5 y^2}{8\pi}\right)^2 - \alpha^2~,
\ee
where $y= N_6 g_s g^2/2\pi$. This then shows that the full solution
\be\label{fundsol}
K = 1 + \f{N_{5} N_6g_s}{32\pi^3}\int_{-\infty}^{\infty} \lambda^2 \Gamma(\tfrac{N_6g_s|\lambda|}{8\pi})\e^{-|\lambda| r+i\lambda z}U(1 + \tfrac{N_6 |\lambda| g_s}{8\pi},2,2 |\lambda| r)\dd \lambda~,
\ee
which describes an intersection of NS5 and D6 branes has an AdS$_7$ space as its ``near-horizon'' geometry.

We can construct even more general solutions with continuous NS5 charge distribution $\sigma_\text{NS5}$ on the $z$-axis. To do this we have to modify the right hand side of equation \eqref{masslessKeq} to:
\be\label{masslesswithsource}
-\triangle_3 K - S(r)\partial_z^2 K = \sigma_\text{NS5}(z)~.
\ee
Since we have already given the solution for which $\sigma_\text{NS5}(z)$ is a delta function in \eqref{fundsol}, we already know the fundamental solution, or Green's function, for the operator $-\triangle_3 - S(r)\partial_z^2$. The homogeneous problem at hand is linear and thus we can use the standard theory of Green's functions to write the solution to the inhomogeneous equation \eqref{masslesswithsource}. The solution is given by convolution of the Green's function with $\sigma_\text{NS5}(z)$. Fourier transform maps convolution to simple multiplication and so the solution is 
\begin{equation}\label{masslesssolutionwithdens}
\begin{split}
K(r,z) =  1 + \tfrac{N_6g_s}{32\pi^3}\int_{-\infty}^{\infty}\lambda^2 \Gamma(\tfrac{N_6g_s|\lambda|}{8\pi})U(1 + \tfrac{N_6 |\lambda| g_s}{8\pi},2,2 r |\lambda|)\hat{\sigma}_\text{NS5}(\lambda)\e^{-|\lambda| r+i\lambda z}\dd \lambda~,
\end{split}
\end{equation}
where $\hat{\sigma}_\text{NS5}$ is the Fourier transform of $\sigma_\text{NS5}$
\begin{equation}
\sigma_\text{NS5}(z)= \f{1}{\sqrt{2\pi}}\int_{-\infty}^{\infty} \hat{\sigma}_\text{NS5}(\lambda)\e^{i\lambda z}\dd \lambda\;.
\end{equation}
In the ``near-horizon'' limit $r\to 0$ we recover the solution \eqref{masslessKasconv} obtained in previous section.

We have thus illustrated how one can construct explicit solutions of type IIA supergravity (with $M=0$) which are asymptotically flat, describe the NS5-D6 brane intersection of interest and have a ``near-horizon'' $AdS_7$ limit. Ideally we would like to be able to do the same for the more general NS5-D6-D8 brane intersection system. However this problem is much more difficult. The cause of trouble are as usual the D8-branes. Due to their presence we have $M\neq 0$ and thus we cannot hope for an asymptotically flat region of space-time. In addition for $M\neq 0$ the BPS equations in \eqref{imamuraeq} can be combined into a single non-linear equation for the function $S$
\be\label{massiveimaeq}
\triangle_3 S + \f12 \partial_z^2(S^2) = 0~.
\ee
This is a non-linear PDE for which we were not able to find the general solution. We found a particular solution of this equation in \eqref{eqSMnot0}-\eqref{eqrznot0} by uplifting the seven-dimensional domain-wall background. However due to the non-linear nature of the problem we cannot use this solution as a seed to construct more general solutions by superposition.

\section{Conclusions}
\label{sec:conclusions}

The three main results of our work can be summarized as follows. First, after carefully studying the BPS equations of massive type IIA supergravity which describe NS5-D6-D8-brane intersections we were able to recover the plethora of $AdS_7$ vacua classified in \cite{Apruzzi:2013yva}. This is a satisfying result and provides additional strong evidence that the supergravity $AdS_7$ solutions of \cite{Apruzzi:2013yva} are indeed dual to the six-dimensional $\mathcal{N}=(1,0)$ SCFT studied in \cite{Gaiotto:2014lca}. Furthermore we utilized a consistent truncation of massive type IIA supergravity to the minimal seven-dimensional gauged supergravity to construct an explicit analytic supersymmetric domain wall solution. This supergravity background can be interpreted holographically as describing a supersymmetric RG flow on the tensor branch of the six-dimensional theory. The flow is triggered by the scalar operator of dimension four which resides in energy-momentum tensor multiplet. Finally, we employed the linear structure of the BPS equations of type IIA supergravity with vanishing Romans mass to construct an explicit supergravity solution which describes the configuration of NS5- and D6-branes schematically presented in Figure \ref{intersection}. Our results lead to many open questions both in supergravity and in field theory and we summarize some of the more pressing ones below.

An important open problem in supergravity is how to find an explicit solution similar to the one in Section \ref{sec:masslessflatsol} which describes the brane intersection of NS5- and D6-branes in the presence of non-vanishing D8-brane charge, $M\neq 0$ (and with the possible addition of O6-planes). This solution should admit a ``near-horizon'' limit in which one recovers the analytic $AdS_7$ solutions presented in Section \ref{subsec:AdS7}. This problem is challenging for at least two reasons. The BPS equations in \eqref{imamuraeq} with $M\neq 0$ are non-linear and one cannot readily find explicit solutions. In addition, due to the presence of D8-branes, one should not expect the background of interest to be asymptotic to flat space and it is a priori not clear what is the correct asymptotic behavior far away from the $AdS_7$ region.

The domain wall supergravity solutions discussed in Section \ref{sec:7dflows} are certainly interesting holographically, however they present a challenge for supergravity. While we have argued that the solutions with $v>0$ are physical and should be dual to a locus on the tensor branch of the six-dimensional SCFT, they are singular in both seven- and ten-dimensional supergravity. It is crucial to understand how to resolve this singularity since this has the potential to teach us interesting lessons about holography as well as about the mechanisms of singularity resolution in string theory. One possible resolution is that the smeared brane densities found in Section \ref{sec:7dflows} localize to branes distributed on a line segment. Such dynamics was observed in the case of smeared NS5-branes where world-sheet instantons lead to clumping of the branes \cite{Tong:2002rq}. An alternative possibility is suggested by the fact that the holographic RG flow at hand preserve eight real supercharges and look similar in spirit to the one of the four-dimensional $\mathcal{N}=2^*$ gauge theory which can be thought of as a mass deformation of $\mathcal{N}=4$ SYM and has been studied extensively in holography and string theory in \cite{Johnson:1999qt,Buchel:2000cn,Evans:2000ct,Pilch:2000ue,Polchinski:2000uf}. It will certainly be very interesting to settle this question.

The vev deformation described holographically by the supergravity domain wall solutions in Section \ref{sec:7dflows} is clearly universal and calls for a better field theory understanding. The scalar operator which drives the flow belongs to the energy-momentum multiplet in the six-dimensional $(1,0)$ SCFT. The supergravity solution suggests that all such SCFTs with holographic duals exhibit this supersymmetric RG flow on their tensor branch. It is certainly desirable to have a field theory understanding of this universal behavior. It will also be interesting to establish a connection between this RG flow on the tensor branch and the field theory and geometric results for similar RG flows in \cite{Intriligator:2014eaa} and \cite{Heckman:2015ola}. 

Finally it should be noted that the six-dimensional SCFTs dual to the $AdS_7$ vacua discussed in Section \ref{subsec:AdS7} admit twisted compactifications to two-, three-, and four-dimensional interacting CFTs with various amounts of supersymmetry \cite{Apruzzi:2015wna,Apruzzi:2015zna,Rota:2015aoa}. It is natural to expect that these lower-dimensional supersymmetric CFTs will in turn have non-trivial vacuum moduli spaces. It will be very interesting to understand whether the ``universal'' tensor branch flow of the 6d theory ``descends'' to some interesting RG flow in the lower-dimensional theory.

We hope that further research will elucidate some of these interesting questions.

\section*{Acknowledgements}

We would like to thank Thomas Van Riet for early collaboration on deriving some of the results discussed in this work and for numerous useful discussions. In addition we are grateful to Marco Baggio, Adam Bzowski, Edoardo Lauria, and Alessandro Tomasiello for useful discussions. The work of NB is supported in part by the starting grant BOF/STG/14/032 from KU Leuven and by an Odysseus grant G0F9516N from the FWO. GD is supported by the Swedish Research Council (VR). FFG is supported in part by the John Templeton Foundation Grant 48222 and by the FWO Odysseus grant G0E5214N. BT is funded by an FWO PhD fellowship. NB, FFG and BT are also supported by the KU Lueven C1 grant ZKD1118 C16/16/005, by the Belgian Federal Science Policy Office through the Inter-University Attraction Pole P7/37, and by the COST Action MP1210 The String Theory Universe.

\appendix
\section{Conventions and notation}
\label{notation}

The action of massive type IIA supergravity in string frame is
\begin{equation}\label{action}
\begin{split}
S =& \f{1}{16\pi G_{10}} \int d^{10}x \sqrt{-g_{10}}\left\{ \e^{-2\phi}\left[R_{10} + 4|\dd \phi|^2 - \f12 |H|^2\right]\right.\\
&\left. - \f12 M^2 - \f12 |F_2|^2 - \f12 |F_4|^2 \right\} +\text{CS-terms}~,
\end{split}
\end{equation}
where $g_{\mu\nu}$ is the ten-dimensional metric in string frame\footnote{To convert to Einstein frame one should use the relation $g_{\mu\nu} = e^{\phi/2}g_{\mu\nu}^{(E)}$, where $\phi$ is the dilaton and $g_{\mu\nu}^{(E)}$ is the metric in Einstein frame. } with Ricci tensor $R_{10}$ in mostly plus conventions and $g_{10}$ is its determinant. The dilaton is denoted by $\phi$, $M$ is the Romans mass \cite{Romans:1985tz}, the three-form field strength is $H=\dd B$ and the RR fields are $F_2$ and $F_4$. We have suppressed the CS terms that ensure the correct equations of motion for $F_2$ and $F_4$. The Bianchi identities are 
\be
\dd H = \dd F_4 - H\w F_2 = \dd F_2 - MH = 0\;.
\ee
We work in string units with
\be
2\pi l_s = 1\;,
\ee
which implies
\be
16\pi G_{10} = \f{1}{2\pi}\;,
\ee
and the string coupling $g_s$ is absorbed in $\e^\phi$. This gives particularly simple quantization conditions for the fluxes \cite{Polchinski:1998rr}, 
namely
\be
M\;,\quad\int H\;,\quad\int \left(F_2-M B\right),\quad \int \left(F_4-B\w F_2+ \f12 M B\w B \right) \in \mathbb{Z}~.
\ee
%

\section{General AdS$_7$ solutions of type IIA supergravity}
\label{app:AdS7}

The general type IIA supergravity solution corresponding to the brane intersection of interest is given in \eqref{ansatz4} and is obtained by solving the system of equations in \eqref{imamuraeq}. These equations are also equivalent to 
\bea
\partial_z L_S &=& Mg_s L_K~,\nonumber\\
\partial_r L_S + M g_s T &=& 0~,\label{Imamuraeqalt}\\
\partial_r L_K + \partial_z T &=& 0~,\nonumber
\eea
where we have defined
\bea
L_S &\equiv (2+2r\partial_r + z\partial_z) S~,\nonumber\\
L_K &\equiv (3+2r\partial_r + z\partial_z) K~,\label{Ldefs}\\
T &\equiv -z\partial_r K + 2r \partial_z(KS)~.\nonumber
\eea
We will now show that $AdS_7$ solutions of the ten-dimensional theory necessarily obey the equations
\be
L_S=L_K = T=0~.
\ee
Furthermore, all AdS$_7$ solutions of the original system can be found in this way.

To find $AdS_7$ within our general Ansatz \eqref{ansatz1}-\eqref{ansatz4} we have to impose that the metric and all background fields are invariant under the isometries of $AdS_7$. To implement this we change coordinates from $(z,r)$ to $(\rho,\alpha)$ where $\rho$ is the radial coordinate of $AdS_7$ and $\alpha$ is a coordinate on the internal space. The metric and three form field strength take the following form 
\begin{equation}
\begin{split}
\dd s^2 =& S^{-\f12} \dd s_6^2 +K[S^{-\f12}(\partial_\rho z)^2+S^{\f12}(\partial_\rho r)^2 ] \dd \rho^2 
\\
&+ K[S^{-\f12}\partial_\rho z\partial_\alpha z+S^{\f12}\partial_\rho r \partial_\alpha r]\dd \rho \dd \alpha
\\
&+ K[S^{-\f12}(\partial_\alpha z)^2+S^{\f12}(\partial_\alpha r)^2] \dd \alpha^2+KS^{\f12}r(\rho,\alpha)^2 \dd \Omega_2^2 \, ,
\\
H =& -r^2 [ \partial_r K \partial_\rho z -\partial_z(KS) \partial_\rho r]\dd \rho \w \Omega_2 
\\
&\, \, -r^2 [ \partial_r K \partial_\alpha z -\partial_z(KS) \partial_\alpha r]\dd \alpha \w \Omega_2 \, .
\end{split}
\end{equation}
Invariance under the isometries of $AdS_7$ requires that the warp factor in front of the $AdS_7$ part of the metric can only depend on the internal coordinate $\alpha$. In addition the three form field strength should only have legs in the internal space. We work with the following metric on $AdS_7$
\be\label{eqn:ads7app}
\dd s_7^2 = \f{1}{(g\rho)^2}\dd\rho^2 + (g\rho)\dd s_6^2~.
\ee
Imposing invariance under the $SO(6,2)$ isometry group of this space leads to the following relations
\bea
S(\rho,\alpha)&=&(g\rho)^{-2}e^{-4A(\alpha)} \label{eq:AdSwarp} \, ,\\
(g\rho)^{-2} e^{2A(\alpha)} &=&K[S^{-\f12}(\partial_\rho z)^2+S^{\f12}(\partial_\rho r)^2 ] \label{eq:metricRho} \, , \\
0&=&K[S^{-\f12}\partial_\rho z\partial_\alpha z+S^{\f12}\partial_\rho r \partial_\alpha r] \label{eq:metricRhoA} \, ,  \\
P(\alpha)&=&K[S^{-\f12}(\partial_\alpha z)^2+S^{\f12}(\partial_\alpha r)^2] \label{eq:metricA}  \, , \\
Q(\alpha)&=&KS^{\f12}r(\rho,\alpha)^2 \label{eq:metricS2}  \, , \\
0&=&\partial_r K \partial_\rho z -\partial_z(KS) \partial_\rho r \label{eq:Hvanish} \, .
\eea
Here we have defined the warp factor in front of $AdS_7$ metric in \eqref{eqn:ads7app} to be $e^{2A(\alpha)}$ and $P(\alpha)$, $Q(\alpha)$ are so far undetermined functions that only depend on $\alpha$. We furthermore have to impose that the dilaton \eqref{ansatz2} depends only on $\alpha$. This condition, combined with \eqref{eq:AdSwarp}, fixes the $\rho$ dependence of the function $K(\rho,\alpha)$ to be
\be
e^\phi = g_s K^{\f12}S^{-\f34} \quad \rightarrow \quad K=g_s^{-2} (g\rho)^{-3} e^{-6A(\alpha)+2\phi(\alpha)} \, .
\ee
From equation (\ref{eq:metricS2}) we immediately see that 
\be \label{eq:rrhoalpha}
r(\rho,\alpha)=(g\rho)^{2} f_1(\alpha) \, ,
\ee
where $f_1(\alpha)=g_Se^{4A-\phi}Q^{-\f12}$ is a nonconstant function of $\alpha$. Using this in turn allows one to rewrite equation (\ref{eq:metricRho})
\be
(\partial_\rho z)^2 = g_s^2e^{6A(\alpha)-2\phi(\alpha)}-4g^2e^{-4A(\alpha)} f_1(\alpha)^2 \equiv g_1(\alpha)^2 \, ,\label{eq:metricRho2} 
\ee
where $g_1(\alpha)$ is defined to notational brevity. With this at hand we can find the $\rho$ dependence of $z(\rho,\alpha)$ to be
 \be\label{eq:zrhoalpha}
 z=\rho g_1(\alpha)+g_2(\alpha) \, .
 \ee
In \eqref{eq:zrhoalpha} we have allowed for an arbitrary function $g_2(\alpha)$, however it is easy to show that $g_2$ has to be a constant. Indeed, from \eqref{eq:metricRhoA} and \eqref{eq:zrhoalpha} one finds
 \be
 0=g_1(\alpha)\left(\rho g_1(\alpha)'+g_2(\alpha)' \right) +e^{-4A(\alpha)} 2g^2\rho f_1(\alpha) f_1(\alpha)'	\, ,
 \ee
which is only consistent if $g_1(\alpha)'\neq0$ and $g_2(\alpha)'=0$. The shift symmetry in $z$ allows us to safely put $g_2=0$. Combining \eqref{eq:rrhoalpha} and \eqref{eq:zrhoalpha} one then finds
\begin{equation}
\frac{r}{z^2} = \frac{g^2f_1(\alpha)}{g_1(\alpha)^2}\;,
\end{equation}
which in turn implies that $\alpha$ has to be a function of $\frac{r}{z^2}$. Furthermore, equation (\ref{eq:metricRho}) can be used to define $\rho$ implicitly, this yields the following relations
\begin{equation}\label{eq:Coordinates}
\alpha=\alpha(r/z^2) \, , \qquad\qquad
\rho^{-1} = g^3K (z^2+4r^2S) \, . 
\end{equation}
Moreover, using $S\sim \rho^{-2}$ and $K\sim \rho^{-3}$ one can show that
\begin{equation}\label{eq:constraint}
\begin{split}
 L_S\equiv 2S + 2r \partial_r S + z \partial_z S=&0 \, ,\\
  L_K\equiv 3K + 2r\partial_r K + z\partial_z K =&0 \, .
\end{split}
\end{equation}
We still need to impose the condition that the three form field strength have legs only along the internal space. This is is given by (\ref{eq:Hvanish}), which in turn leads to
\be
T\equiv -z\partial_r K + 2r\partial_z(KS) = 0~,
\label{eq:constraint2}
\ee
The only thing left to show is that the conditions in (\ref{eq:metricRhoA}) and (\ref{eq:metricA}) are satisfied. For this we need to invert the following Jacobian
\be
\frac{\partial(\rho,\alpha)}{\partial(z,r)}=\left( \begin{matrix}	\partial_z\rho & & \partial_r \rho \\ & & \\ -2\frac{r}{z^3}\alpha' & & \frac{1}{z^2}\alpha'	\end{matrix}\right) \, , \label{eq:RhoAzr}
\ee
where $\alpha'$ denotes the derivative of $\alpha$ with respect to $r/z^2$ and one can use the relations
\begin{equation}
\begin{split}
\partial_z\rho  =& -\rho^2 \left( \rho^{-1} \frac{\partial_zK}{K} +g^3K(2z+4r^2\partial_zS) \right)\;, \\
\partial_r \rho =& -\rho^2 \left( \rho^{-1} \frac{\partial_rK}{K}+g^3K(8rS+4r^2\partial_rS)\right) \,.
\end{split}
\end{equation}
Using (\ref{eq:constraint}) one finds that
\be
\frac{\partial(z,r)}{\partial(\rho,\alpha)}=\frac{z^3\rho}{\alpha' }\left( \begin{matrix}	\frac{1}{z^2}\alpha' & & -\partial_r \rho \\ & & \\ 2\frac{r}{z^3}\alpha' & & \partial_z\rho\end{matrix}\right) \,. \label{eq ZRrhoA}
\ee
This equation can then be used to show the validity of (\ref{eq:metricRhoA}, \ref{eq:metricA}). This concludes the prove that all supersymmetric AdS$_7$ solutions of type IIA supergravity should obey the constraints in (\ref{eq:constraint}, \ref{eq:constraint2}). In Section \ref{subsec:AdS7} we show how to explicitly solve these constraints and find all of these $AdS_7$ solutions analytically.

\section{Comparison to the results in \cite{Apruzzi:2013yva}}
\label{AFRT}

In this appendix we show that the general system of equations for supersymmetric $AdS_7$ backgrounds of massive type IIA supergravity derived in \cite{Apruzzi:2013yva} is solved by the background in \eqref{metricfinal} together with the equations in \eqref{AdsODE}. In order to match the conventions used in this paper we flip the signs of $M$ and $F_2$ appearing in the system of \cite{Apruzzi:2013yva}. We must also take $g=2$ since in \cite{Apruzzi:2013yva} the authors fix the radius of $AdS_7$ to be $L=2/g=1$. The $AdS_7$ solutions of \cite{Apruzzi:2013yva} (in string frame) are given by
\begin{equation}
\begin{split}
\dd s^2 &=  \e^{2A}\left(\dd s_{\text{AdS}_7}^2 + \dd s_{M_3}^2\right)~,\\
\dd s_{M_3}^2 &= (1-x^2)\left(\f{16}{(4x+M\e^{A+\phi})^2}\dd A^2 + \f{1}{16}\dd\Omega_2^2\right)~,\\
F_2 &= \f{\sqrt{1-x^2}}{16}\e^{A-\phi}(xM\e^{A+\phi}+4)\, \Omega_2 ~,\\
H &= -\f14 \e^{2A}(1-x^2)^{\frac{3}{2}} \f{6-xM\e^{A+\phi}}{4x+M\e^{A+\phi}}\dd A\w \Omega_2 ~.
\end{split}
\end{equation}
This constitutes a supersymmetric background of massive type IIA supergravity provided that the dilaton $\phi(A)$ and the function $x(A)$ satisfy the pair of coupled ordinary differential equations
\begin{equation}
\begin{split}
\partial_A \phi &= 5-2x^2+\f{8x(x^2-1)}{4x+M\e^{A+\phi}}~,\\
\partial_A x &= 2(x^2-1)\f{4-xM\e^{A+\phi}}{4x+M\e^{A+\phi}}~.
\end{split}
\end{equation}
We find that the background above agrees with our expression in \eqref{metricfinal} if we set
\begin{equation}
\begin{split}
\e^{2A} &= \sqrt{\f{\beta}{y}}~,\\
x^2 &= \f{\alpha^2}{\alpha^2+4y\beta}~,\\
\dd A &= -\f1{8y}\left(\f{\alpha}{\beta} - \f{M g_s}{y}\right)~\dd \alpha~.
\end{split}
\end{equation}
Furthermore, both differential equations are solved provided the equations in \eqref{AdsODE} are obeyed.

\bibliographystyle{utphys}
\bibliography{refs}

\providecommand{\href}[2]{#2}\begingroup\raggedright\begin{thebibliography}{10}

\bibitem{Apruzzi:2013yva}
F.~Apruzzi, M.~Fazzi, D.~Rosa, and A.~Tomasiello, ``{All AdS$_7$ solutions of
  type II supergravity},''
  \href{http://dx.doi.org/10.1007/JHEP04(2014)064}{{\em JHEP} {\bfseries 04}
  (2014) 064},
\href{http://arxiv.org/abs/1309.2949}{{\ttfamily arXiv:1309.2949 [hep-th]}}.

\bibitem{Seiberg:1996qx}
N.~Seiberg, ``{Nontrivial fixed points of the renormalization group in
  six-dimensions},''
  \href{http://dx.doi.org/10.1016/S0370-2693(96)01424-4}{{\em Phys. Lett.}
  {\bfseries B390} (1997) 169--171},
\href{http://arxiv.org/abs/hep-th/9609161}{{\ttfamily arXiv:hep-th/9609161
  [hep-th]}}.

\bibitem{Brunner:1997gf}
I.~Brunner and A.~Karch, ``{Branes at orbifolds versus Hanany Witten in
  six-dimensions},''
  \href{http://dx.doi.org/10.1088/1126-6708/1998/03/003}{{\em JHEP} {\bfseries
  03} (1998) 003},
\href{http://arxiv.org/abs/hep-th/9712143}{{\ttfamily arXiv:hep-th/9712143
  [hep-th]}}.

\bibitem{Hanany:1997gh}
A.~Hanany and A.~Zaffaroni, ``{Branes and six-dimensional supersymmetric
  theories},'' \href{http://dx.doi.org/10.1016/S0550-3213(98)00355-1}{{\em
  Nucl. Phys.} {\bfseries B529} (1998) 180--206},
\href{http://arxiv.org/abs/hep-th/9712145}{{\ttfamily arXiv:hep-th/9712145
  [hep-th]}}.

\bibitem{Heckman:2013pva}
J.~J. Heckman, D.~R. Morrison, and C.~Vafa, ``{On the Classification of 6D
  SCFTs and Generalized ADE Orbifolds},''
  \href{http://dx.doi.org/10.1007/JHEP06(2015)017,
  10.1007/JHEP05(2014)028}{{\em JHEP} {\bfseries 05} (2014) 028},
  \href{http://arxiv.org/abs/1312.5746}{{\ttfamily arXiv:1312.5746 [hep-th]}}.
[Erratum: JHEP06,017(2015)].

\bibitem{DelZotto:2014hpa}
M.~Del~Zotto, J.~J. Heckman, A.~Tomasiello, and C.~Vafa, ``{6d Conformal
  Matter},'' \href{http://dx.doi.org/10.1007/JHEP02(2015)054}{{\em JHEP}
  {\bfseries 02} (2015) 054},
\href{http://arxiv.org/abs/1407.6359}{{\ttfamily arXiv:1407.6359 [hep-th]}}.

\bibitem{Heckman:2015bfa}
J.~J. Heckman, D.~R. Morrison, T.~Rudelius, and C.~Vafa, ``{Atomic
  Classification of 6D SCFTs},''
  \href{http://dx.doi.org/10.1002/prop.201500024}{{\em Fortsch. Phys.}
  {\bfseries 63} (2015) 468--530},
\href{http://arxiv.org/abs/1502.05405}{{\ttfamily arXiv:1502.05405 [hep-th]}}.

\bibitem{Gaiotto:2014lca}
D.~Gaiotto and A.~Tomasiello, ``{Holography for (1,0) theories in six
  dimensions},'' \href{http://dx.doi.org/10.1007/JHEP12(2014)003}{{\em JHEP}
  {\bfseries 12} (2014) 003},
\href{http://arxiv.org/abs/1404.0711}{{\ttfamily arXiv:1404.0711 [hep-th]}}.

\bibitem{Apruzzi:2015zna}
F.~Apruzzi, M.~Fazzi, A.~Passias, and A.~Tomasiello, ``{Supersymmetric
  AdS$_{5}$ solutions of massive IIA supergravity},''
  \href{http://dx.doi.org/10.1007/JHEP06(2015)195}{{\em JHEP} {\bfseries 06}
  (2015) 195},
\href{http://arxiv.org/abs/1502.06620}{{\ttfamily arXiv:1502.06620 [hep-th]}}.

\bibitem{Rota:2015aoa}
A.~Rota and A.~Tomasiello, ``{AdS$_{4}$ compactifications of AdS$_{7}$
  solutions in type II supergravity},''
  \href{http://dx.doi.org/10.1007/JHEP07(2015)076}{{\em JHEP} {\bfseries 07}
  (2015) 076},
\href{http://arxiv.org/abs/1502.06622}{{\ttfamily arXiv:1502.06622 [hep-th]}}.

\bibitem{Apruzzi:2015wna}
F.~Apruzzi, M.~Fazzi, A.~Passias, A.~Rota, and A.~Tomasiello,
  ``{Six-Dimensional Superconformal Theories and their Compactifications from
  Type IIA Supergravity},''
  \href{http://dx.doi.org/10.1103/PhysRevLett.115.061601}{{\em Phys. Rev.
  Lett.} {\bfseries 115} no.~6, (2015) 061601},
\href{http://arxiv.org/abs/1502.06616}{{\ttfamily arXiv:1502.06616 [hep-th]}}.

\bibitem{Ohmori:2014kda}
K.~Ohmori, H.~Shimizu, Y.~Tachikawa, and K.~Yonekura, ``{Anomaly polynomial of
  general 6d SCFTs},'' \href{http://dx.doi.org/10.1093/ptep/ptu140}{{\em PTEP}
  {\bfseries 2014} no.~10, (2014) 103B07},
\href{http://arxiv.org/abs/1408.5572}{{\ttfamily arXiv:1408.5572 [hep-th]}}.

\bibitem{Tomasiello:2016wfy}
A.~Tomasiello, ``{Higher-dimensional gauge theories from string theory},''
\href{http://dx.doi.org/10.1002/prop.201500089}{{\em Fortsch. Phys.} {\bfseries
  64} (2016) 303--316}.

\bibitem{Bhardwaj:2015xxa}
L.~Bhardwaj, ``{Classification of 6d $ \mathcal{N}=\left(1,0\right) $ gauge
  theories},'' \href{http://dx.doi.org/10.1007/JHEP11(2015)002}{{\em JHEP}
  {\bfseries 11} (2015) 002},
\href{http://arxiv.org/abs/1502.06594}{{\ttfamily arXiv:1502.06594 [hep-th]}}.

\bibitem{Hanany:1996ie}
A.~Hanany and E.~Witten, ``{Type IIB superstrings, BPS monopoles, and
  three-dimensional gauge dynamics},''
  \href{http://dx.doi.org/10.1016/S0550-3213(97)00157-0,
  10.1016/S0550-3213(97)80030-2}{{\em Nucl. Phys.} {\bfseries B492} (1997)
  152--190},
\href{http://arxiv.org/abs/hep-th/9611230}{{\ttfamily arXiv:hep-th/9611230
  [hep-th]}}.

\bibitem{Witten:1997sc}
E.~Witten, ``{Solutions of four-dimensional field theories via M theory},''
  \href{http://dx.doi.org/10.1016/S0550-3213(97)00416-1}{{\em Nucl. Phys.}
  {\bfseries B500} (1997) 3--42},
\href{http://arxiv.org/abs/hep-th/9703166}{{\ttfamily arXiv:hep-th/9703166
  [hep-th]}}.

\bibitem{Cremonesi:2015bld}
S.~Cremonesi and A.~Tomasiello, ``{6d holographic anomaly match as a continuum
  limit},'' \href{http://dx.doi.org/10.1007/JHEP05(2016)031}{{\em JHEP}
  {\bfseries 05} (2016) 031},
\href{http://arxiv.org/abs/1512.02225}{{\ttfamily arXiv:1512.02225 [hep-th]}}.

\bibitem{Maldacena:1997re}
J.~M. Maldacena, ``{The Large N limit of superconformal field theories and
  supergravity},'' \href{http://dx.doi.org/10.1023/A:1026654312961}{{\em Int.
  J. Theor. Phys.} {\bfseries 38} (1999) 1113--1133},
  \href{http://arxiv.org/abs/hep-th/9711200}{{\ttfamily arXiv:hep-th/9711200
  [hep-th]}}.
[Adv. Theor. Math. Phys.2,231(1998)].

\bibitem{Imamura:2001cr}
Y.~Imamura, ``{1/4 BPS solutions in massive IIA supergravity},''
  \href{http://dx.doi.org/10.1143/PTP.106.653}{{\em Prog. Theor. Phys.}
  {\bfseries 106} (2001) 653--670},
\href{http://arxiv.org/abs/hep-th/0105263}{{\ttfamily arXiv:hep-th/0105263
  [hep-th]}}.

\bibitem{Passias:2015gya}
A.~Passias, A.~Rota, and A.~Tomasiello, ``{Universal consistent truncation for
  6d/7d gauge/gravity duals},''
  \href{http://dx.doi.org/10.1007/JHEP10(2015)187}{{\em JHEP} {\bfseries 10}
  (2015) 187},
\href{http://arxiv.org/abs/1506.05462}{{\ttfamily arXiv:1506.05462 [hep-th]}}.

\bibitem{Cordova:2016xhm}
C.~Cordova, T.~T. Dumitrescu, and K.~Intriligator, ``{Deformations of
  Superconformal Theories},''
\href{http://arxiv.org/abs/1602.01217}{{\ttfamily arXiv:1602.01217 [hep-th]}}.

\bibitem{Macpherson:2016xwk}
N.~T. Macpherson and A.~Tomasiello, ``{Minimal flux Minkowski
  classification},''
\href{http://arxiv.org/abs/1612.06885}{{\ttfamily arXiv:1612.06885 [hep-th]}}.

\bibitem{Louis:2015mka}
J.~Louis and S.~Lüst, ``{Supersymmetric AdS$_{7}$ backgrounds in half-maximal
  supergravity and marginal operators of (1, 0) SCFTs},''
  \href{http://dx.doi.org/10.1007/JHEP10(2015)120}{{\em JHEP} {\bfseries 10}
  (2015) 120},
\href{http://arxiv.org/abs/1506.08040}{{\ttfamily arXiv:1506.08040 [hep-th]}}.

\bibitem{Buican:2016hpb}
M.~Buican, J.~Hayling, and C.~Papageorgakis, ``{Aspects of Superconformal
  Multiplets in D>4},''
\href{http://arxiv.org/abs/1606.00810}{{\ttfamily arXiv:1606.00810 [hep-th]}}.

\bibitem{Cordova:2015fha}
C.~Cordova, T.~T. Dumitrescu, and K.~Intriligator, ``{Anomalies,
  Renormalization Group Flows, and the a-Theorem in Six-Dimensional (1,0)
  Theories},''
\href{http://arxiv.org/abs/1506.03807}{{\ttfamily arXiv:1506.03807 [hep-th]}}.

\bibitem{Romans:1985tz}
L.~J. Romans, ``{Massive N=2a Supergravity in Ten-Dimensions},''
\href{http://dx.doi.org/10.1016/0370-2693(86)90375-8}{{\em Phys. Lett.}
  {\bfseries B169} (1986) 374}.

\bibitem{Janssen:1999sa}
B.~Janssen, P.~Meessen, and T.~Ortin, ``{The D8-brane tied up: String and brane
  solutions in massive type IIA supergravity},''
  \href{http://dx.doi.org/10.1016/S0370-2693(99)00315-9}{{\em Phys. Lett.}
  {\bfseries B453} (1999) 229--236},
\href{http://arxiv.org/abs/hep-th/9901078}{{\ttfamily arXiv:hep-th/9901078
  [hep-th]}}.

\bibitem{Blaback:2011pn}
J.~Blaback, U.~H. Danielsson, D.~Junghans, T.~Van~Riet, T.~Wrase, and
  M.~Zagermann, ``{(Anti-)Brane backreaction beyond perturbation theory},''
  \href{http://dx.doi.org/10.1007/JHEP02(2012)025}{{\em JHEP} {\bfseries 02}
  (2012) 025},
\href{http://arxiv.org/abs/1111.2605}{{\ttfamily arXiv:1111.2605 [hep-th]}}.

\bibitem{Myers:1999ps}
R.~C. Myers, ``{Dielectric branes},''
  \href{http://dx.doi.org/10.1088/1126-6708/1999/12/022}{{\em JHEP} {\bfseries
  12} (1999) 022},
\href{http://arxiv.org/abs/hep-th/9910053}{{\ttfamily arXiv:hep-th/9910053
  [hep-th]}}.

\bibitem{Junghans:2014wda}
D.~Junghans, D.~Schmidt, and M.~Zagermann, ``{Curvature-induced Resolution of
  Anti-brane Singularities},''
  \href{http://dx.doi.org/10.1007/JHEP10(2014)034}{{\em JHEP} {\bfseries 10}
  (2014) 34},
\href{http://arxiv.org/abs/1402.6040}{{\ttfamily arXiv:1402.6040 [hep-th]}}.

\bibitem{Apruzzi:2016rny}
F.~Apruzzi, G.~Dibitetto, and L.~Tizzano, ``{A new 6d fixed point from
  holography},''
\href{http://arxiv.org/abs/1603.06576}{{\ttfamily arXiv:1603.06576 [hep-th]}}.

\bibitem{Kraus:1998hv}
P.~Kraus, F.~Larsen, and S.~P. Trivedi, ``{The Coulomb branch of gauge theory
  from rotating branes},''
  \href{http://dx.doi.org/10.1088/1126-6708/1999/03/003}{{\em JHEP} {\bfseries
  03} (1999) 003},
\href{http://arxiv.org/abs/hep-th/9811120}{{\ttfamily arXiv:hep-th/9811120
  [hep-th]}}.

\bibitem{Freedman:1999gk}
D.~Z. Freedman, S.~S. Gubser, K.~Pilch, and N.~P. Warner, ``{Continuous
  distributions of D3-branes and gauged supergravity},''
  \href{http://dx.doi.org/10.1088/1126-6708/2000/07/038}{{\em JHEP} {\bfseries
  07} (2000) 038},
\href{http://arxiv.org/abs/hep-th/9906194}{{\ttfamily arXiv:hep-th/9906194
  [hep-th]}}.

\bibitem{Cvetic:1999xx}
M.~Cvetic, S.~S. Gubser, H.~Lu, and C.~N. Pope, ``{Symmetric potentials of
  gauged supergravities in diverse dimensions and Coulomb branch of gauge
  theories},'' \href{http://dx.doi.org/10.1103/PhysRevD.62.086003}{{\em Phys.
  Rev.} {\bfseries D62} (2000) 086003},
\href{http://arxiv.org/abs/hep-th/9909121}{{\ttfamily arXiv:hep-th/9909121
  [hep-th]}}.

\bibitem{Bakas:1999fa}
I.~Bakas, A.~Brandhuber, and K.~Sfetsos, ``{Domain walls of gauged
  supergravity, M-branes, and algebraic curves},'' {\em Adv. Theor. Math.
  Phys.} {\bfseries 3} (1999) 1657--1719,
\href{http://arxiv.org/abs/hep-th/9912132}{{\ttfamily arXiv:hep-th/9912132
  [hep-th]}}.

\bibitem{Townsend:1983kk}
P.~K. Townsend and P.~van Nieuwenhuizen, ``{Gauged Seven-Dimensional
  Supergravity},''
\href{http://dx.doi.org/10.1016/0370-2693(83)91230-3}{{\em Phys. Lett.}
  {\bfseries B125} (1983) 41--46}.

\bibitem{Pernici:1984zw}
M.~Pernici, K.~Pilch, P.~van Nieuwenhuizen, and N.~P. Warner, ``{Noncompact
  Gaugings and Critical Points of Maximal Supergravity in Seven-dimensions},''
\href{http://dx.doi.org/10.1016/0550-3213(85)90046-X}{{\em Nucl. Phys.}
  {\bfseries B249} (1985) 381--395}.

\bibitem{Gubser:2000nd}
S.~S. Gubser, ``{Curvature singularities: The Good, the bad, and the naked},''
  {\em Adv. Theor. Math. Phys.} {\bfseries 4} (2000) 679--745,
\href{http://arxiv.org/abs/hep-th/0002160}{{\ttfamily arXiv:hep-th/0002160
  [hep-th]}}.

\bibitem{Maldacena:2000mw}
J.~M. Maldacena and C.~Nunez, ``{Supergravity description of field theories on
  curved manifolds and a no go theorem},''
  \href{http://dx.doi.org/10.1142/S0217751X01003935,
  10.1142/S0217751X01003937}{{\em Int. J. Mod. Phys.} {\bfseries A16} (2001)
  822--855}, \href{http://arxiv.org/abs/hep-th/0007018}{{\ttfamily
  arXiv:hep-th/0007018 [hep-th]}}.
[,182(2000)].

\bibitem{Clarkson:2004eq}
R.~Clarkson, A.~M. Ghezelbash, and R.~B. Mann, ``{New reducible five-brane
  solutions in M-theory},''
  \href{http://dx.doi.org/10.1088/1126-6708/2004/08/025}{{\em JHEP} {\bfseries
  08} (2004) 025},
\href{http://arxiv.org/abs/hep-th/0405148}{{\ttfamily arXiv:hep-th/0405148
  [hep-th]}}.

\bibitem{TEMME198127}
N.~Temme, ``On the expansion of confluent hypergeometric functions in terms of
  bessel functions,''
  \href{http://dx.doi.org/http://dx.doi.org/10.1016/0771-050X(81)90004-8}{{\em
  Journal of Computational and Applied Mathematics} {\bfseries 7} no.~1, (1981)
  27 -- 32}.

\bibitem{Tong:2002rq}
D.~Tong, ``{NS5-branes, T duality and world sheet instantons},''
  \href{http://dx.doi.org/10.1088/1126-6708/2002/07/013}{{\em JHEP} {\bfseries
  07} (2002) 013},
\href{http://arxiv.org/abs/hep-th/0204186}{{\ttfamily arXiv:hep-th/0204186
  [hep-th]}}.

\bibitem{Johnson:1999qt}
C.~V. Johnson, A.~W. Peet, and J.~Polchinski, ``{Gauge theory and the excision
  of repulson singularities},''
  \href{http://dx.doi.org/10.1103/PhysRevD.61.086001}{{\em Phys. Rev.}
  {\bfseries D61} (2000) 086001},
\href{http://arxiv.org/abs/hep-th/9911161}{{\ttfamily arXiv:hep-th/9911161
  [hep-th]}}.

\bibitem{Buchel:2000cn}
A.~Buchel, A.~W. Peet, and J.~Polchinski, ``{Gauge dual and noncommutative
  extension of an N=2 supergravity solution},''
  \href{http://dx.doi.org/10.1103/PhysRevD.63.044009}{{\em Phys. Rev.}
  {\bfseries D63} (2001) 044009},
\href{http://arxiv.org/abs/hep-th/0008076}{{\ttfamily arXiv:hep-th/0008076
  [hep-th]}}.

\bibitem{Evans:2000ct}
N.~J. Evans, C.~V. Johnson, and M.~Petrini, ``{The Enhancon and N=2 gauge
  theory: Gravity RG flows},''
  \href{http://dx.doi.org/10.1088/1126-6708/2000/10/022}{{\em JHEP} {\bfseries
  10} (2000) 022},
\href{http://arxiv.org/abs/hep-th/0008081}{{\ttfamily arXiv:hep-th/0008081
  [hep-th]}}.

\bibitem{Pilch:2000ue}
K.~Pilch and N.~P. Warner, ``{N=2 supersymmetric RG flows and the IIB
  dilaton},'' \href{http://dx.doi.org/10.1016/S0550-3213(00)00656-8}{{\em Nucl.
  Phys.} {\bfseries B594} (2001) 209--228},
\href{http://arxiv.org/abs/hep-th/0004063}{{\ttfamily arXiv:hep-th/0004063
  [hep-th]}}.

\bibitem{Polchinski:2000uf}
J.~Polchinski and M.~J. Strassler, ``{The String dual of a confining
  four-dimensional gauge theory},''
\href{http://arxiv.org/abs/hep-th/0003136}{{\ttfamily arXiv:hep-th/0003136
  [hep-th]}}.

\bibitem{Intriligator:2014eaa}
K.~Intriligator, ``{6d, $ \mathcal{N}=\left(1,\;0\right) $ Coulomb branch
  anomaly matching},'' \href{http://dx.doi.org/10.1007/JHEP10(2014)162}{{\em
  JHEP} {\bfseries 10} (2014) 162},
\href{http://arxiv.org/abs/1408.6745}{{\ttfamily arXiv:1408.6745 [hep-th]}}.

\bibitem{Heckman:2015ola}
J.~J. Heckman, D.~R. Morrison, T.~Rudelius, and C.~Vafa, ``{Geometry of 6D RG
  Flows},'' \href{http://dx.doi.org/10.1007/JHEP09(2015)052}{{\em JHEP}
  {\bfseries 09} (2015) 052},
\href{http://arxiv.org/abs/1505.00009}{{\ttfamily arXiv:1505.00009 [hep-th]}}.

\bibitem{Polchinski:1998rr}
J.~Polchinski, {\em {String theory. Vol. 2: Superstring theory and beyond}}.
\newblock Cambridge University Press,
2007.
\newblock

\end{thebibliography}\endgroup

\end{document}